\pdfoutput=1

\documentclass[10pt,twocolumn]{article}

\usepackage[T1]{fontenc}
\usepackage[utf8]{inputenc}
\usepackage{mathptmx}
\usepackage[a4paper,margin=2cm,columnsep=0.8cm]{geometry}
\usepackage{graphicx}
\usepackage{amsmath}
\usepackage[table]{xcolor}
\newif\ifauthoryear \authoryeartrue
\ifauthoryear
  \usepackage[round,authoryear]{natbib}
  \DeclareRobustCommand{\cite}[1]{\unskip~\citep{#1}}
  \newcommand{\refcite}[1]{\citet{#1}}
\else
  \usepackage[super]{cite}
  \newcommand{\refcite}[1]{ref.~\cite{#1}}
\fi
\usepackage[colorlinks,allcolors=blue!50!black]{hyperref}
\usepackage[british]{babel}
\makeatletter
\newcommand\FloatBarrier{\par\begingroup\let\@elt\relax
  \edef\@tempa{\@botlist\@deferlist\@dbldeferlist}%
  \ifx\@tempa\@empty\else
    \ifx\@fltovf\relax
      \if@firstcolumn\clearpage\else\null\newpage\clearpage\fi
    \else
      \newpage\let\@fltovf\relax\FloatBarrier
    \fi
  \fi\endgroup}
\makeatother
\newcommand{\mref}[1]{Methods, \nameref{#1}}

\graphicspath{{figures/}}

\newcommand{\placefig}[3]{\IfFileExists{figures/#2}%
  {\includegraphics[width=#1]{#2}}%
  {\fbox{\begin{minipage}[c][0.45#1][c]{\dimexpr#1-2\fboxsep-2\fboxrule\relax}
    \centering\small\texttt{\detokenize{figures/#2}}\par\medskip #3
  \end{minipage}}}}

\newif\ifdraft \draftfalse

\usepackage{tikz}
\definecolor{iaink}{HTML}{16202E}
\definecolor{iashape}{HTML}{27347A}
\definecolor{iablue}{HTML}{2166AC}
\definecolor{iared}{HTML}{B2182B}
\newcommand{\iacl}{\fill[iaink] (0,0) circle (0.26);}
\newcommand{\iavd}{\draw[iaink,line width=0.45pt] (0,0) circle (0.26);}
\newcommand{\iaglyph}[3]{\tikz[baseline=-0.55ex,x=1.35ex,y=1.35ex]{#3%
  \fill[#1,rotate around={#2:(1.15,0)}] (1.15,0) ellipse (0.6 and 0.24);}}
\newcommand{\iarad}[1][iashape]{\iaglyph{#1}{0}{\iacl}}
\newcommand{\iatan}[1][iashape]{\iaglyph{#1}{90}{\iacl}}
\newcommand{\iacross}[1][iashape]{\iaglyph{#1}{45}{\iacl}}
\newcommand{\iaradv}[1][iashape]{\iaglyph{#1}{0}{\iavd}}
\newcommand{\iatanv}[1][iashape]{\iaglyph{#1}{90}{\iavd}}
\newcommand{\iaspiral}{\tikz[baseline=-0.6ex,x=1.45ex,y=1.45ex]{\begin{scope}[rotate=30,yscale=0.82]
  \fill[iablue!18] (0,0) circle (0.92);
  \foreach \a in {0,180}{\draw[iablue,line width=0.75pt,line cap=round,rotate=\a]
    plot[domain=0:4.2,samples=50,smooth] ({0.2*exp(0.36*\x)*cos(\x r)},{0.2*exp(0.36*\x)*sin(\x r)});}
  \shade[inner color=iablue!25!white,outer color=iablue] (0,0) circle (0.3);\end{scope}}}
\newcommand{\iaelliptical}{\tikz[baseline=-0.6ex,x=1.25ex,y=1.25ex]{%
  \shade[rotate=30,inner color=iared!35!white,outer color=iared!90!black] (0,0) ellipse (0.95 and 0.65);}}

\title{\textbf{Star-forming disc galaxies align tangentially with the cosmic tidal field}}
\author{
  Calum Murray\textsuperscript{1,*},
  Romain Paviot\textsuperscript{1,2},
  Martin Kilbinger\textsuperscript{1},
  \\
  Cail Daley\textsuperscript{1},
  Ken Ganga\textsuperscript{3} and
  Michael J. Hudson\textsuperscript{4,5,6}
  \\[1ex]
  \small\textsuperscript{1}Universit\'e Paris-Saclay, Universit\'e Paris Cit\'e, CEA, CNRS, AIM, 91191 Gif-sur-Yvette, France\\
  \small\textsuperscript{2}Centre National d'\'Etudes Spatiales -- Centre spatial de Toulouse, 18 avenue Edouard Belin, 31401 Toulouse Cedex 9, France\\
  \small\textsuperscript{3}Universit\'e Paris Cit\'e, CNRS, Astroparticule et Cosmologie, F-75013 Paris, France\\
  \small\textsuperscript{4}Department of Physics and Astronomy, University of Waterloo, 200 University Avenue West, Waterloo, Ontario N2L 3G1, Canada\\
  \small\textsuperscript{5}Waterloo Centre for Astrophysics, University of Waterloo, Waterloo, Ontario N2L 3G1, Canada\\
  \small\textsuperscript{6}Perimeter Institute for Theoretical Physics, 31 Caroline St. North, Waterloo, Ontario N2L 2Y5, Canada\\
  \small\textsuperscript{*}Corresponding author: calumhrmurray@gmail.com, calum.murray@cea.fr
}
\date{}

\begin{document}

\twocolumn[\begin{@twocolumnfalse}
\maketitle
\begin{abstract}
\noindent
The alignments of galaxy shapes, with one another and with the large-scale structure of the Universe, have been a topic of considerable investigation within cosmology and extragalactic astrophysics for the last 60 years. Galaxy alignments are a major systematic for weak gravitational lensing and help us learn about the processes of galaxy formation. Red early-type galaxies are known to show strong radial alignments around other galaxies; however, blue star-forming galaxies have shown little evidence of alignment. We combine SPHEREx spectrophotometry with DESI Legacy Imaging Surveys to build a full-sky catalogue of 71 million galaxies with precise redshifts. With this catalogue we are able to measure alignment signals by splitting galaxies into colour and morphological samples. We detect tangential alignment of blue disc galaxies at $22.5\sigma$ and radial alignment of red spheroidal galaxies at $65\sigma$. By measuring the alignment signal separately around overdensities and underdensities, we show that each of these alignments has the scale and environmental dependence of a linear response to the cosmic tidal field; a response purely quadratic in the Gaussian tidal field, as in classic tidal torque theory is ruled out. 
\end{abstract}
\vspace{2ex}
\end{@twocolumnfalse}]

\begin{figure*}[tbp]
  \centering
  \placefig{\textwidth}{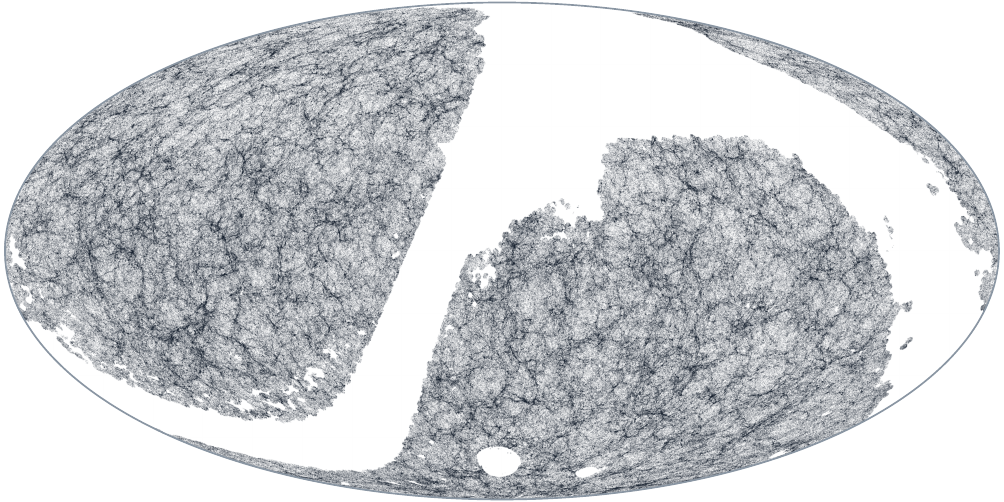}{Full-sky cosmic web, $0.115<z<0.145$}
  \caption{\textbf{The cosmic web}. The black dots are the positions of galaxies from the Legacy-SPHEREx catalogue (LSx) within the redshift range $0.115<z<0.145$, in an equatorial Mollweide projection. The filaments, voids and clusters of the cosmic web are directly visible. An animated version of this figure can be found at \url{https://calumhrmurray.github.io/spherex-sky.html}.}
  \label{fig:cosmicweb}
\end{figure*}

The large-scale alignments of galaxy shapes that we observe in the Universe are thought to be set during the process of galaxy formation. As the process of galaxy formation varies significantly with galaxy type, from spirals and discs to irregulars and spheroids, we should also expect the alignment of galaxy shapes to depend strongly upon galaxy type. Observations of galaxy alignments will help us understand the process of galaxy formation in more detail, and may hold a fossil record of the tidal field at the time of a galaxy's formation.

There are two physical galaxy formation mechanisms which are expected to create large-scale alignments between galaxy shapes. Both are related to the tidal field of the Universe. The tidal field is the differential pull of gravity, and therefore describes the stretching and squeezing produced by the gravitational potential at a given point. A forming galaxy can be stretched directly by the tidal field, or it can be spun up by it through tidal torquing. These lead to the two principal alignment mechanisms, a linear and a quadratic response of the galaxy shape to the tidal field. However, as we explain below, the formation mechanisms and the alignment mechanisms do not map one to one: tidal torquing can also produce a linear response.

For early-type galaxies~\iaelliptical\ the acquisition of an alignment is expected to be relatively simple. As a proto-galaxy forms from gravitational collapse, the cosmic tidal field sets a preferential collapse axis, such that the galaxy's form contains a small additional shear which is proportional to the tidal field\cite{Catelan2001,HirataSeljak2004}. As a result, the projected image of the galaxy has a small linear contribution from the tidal shear. For disc galaxies~\iaspiral\ the situation is significantly more complex. Their alignments are instead expected to be largely determined by the direction of the galaxy's angular momentum\cite{Catelan2001,LeePen2000}. As the proto-galaxy forms, its own inertia tensor may be misaligned with the large-scale tidal field. The tidal field will then torque the proto-galaxy, causing its angular momentum to grow\cite{Peebles1969,Doroshkevich1970,White1984}. As the proto-galaxy inertia also depends upon the tidal field, the resulting alignment of the disc galaxy will respond quadratically to the tidal field.

Considering galaxy shapes projected on the sky, instead of the full three-dimensional bodies, we can approximate the shape of a galaxy as an ellipse. We can write the observed galaxy ellipticity in terms of a contribution from a linear response to the tidal field, $C_1$, a contribution from a quadratic response to the tidal field, $C_2$, and a term from the weighting of the tidal field by the local density field, $C_{1\delta}$, which arises because we can only measure galaxy shapes where there are galaxies. We can sketch the observed galaxy shape field, the average ellipticity of galaxies, $\gamma$, at a position $\mathbf x$, as

\begin{equation}
  \gamma(\mathbf x) = C_1\,s_{\rm lin}(\mathbf x) + C_{1\delta}\,\delta(\mathbf x)\,s_{\rm lin}(\mathbf x) + C_2\,s_{\rm quad}(\mathbf x) ,
\label{eq:response}
\end{equation}

where $\delta$  is the density field, $s_{\rm lin}$ is the tidal field, and $s_{\rm quad}$ is quadratic in the tidal field. Therefore the amplitudes of $C_1$, $C_{1\delta}$ and $C_2$ will determine the shape of the galaxy alignment field. The full form of this equation is given in \mref{sec:model} (Eq.~\eqref{eq:tatt}).

\begin{figure}[htbp]
  \centering
  \placefig{\columnwidth}{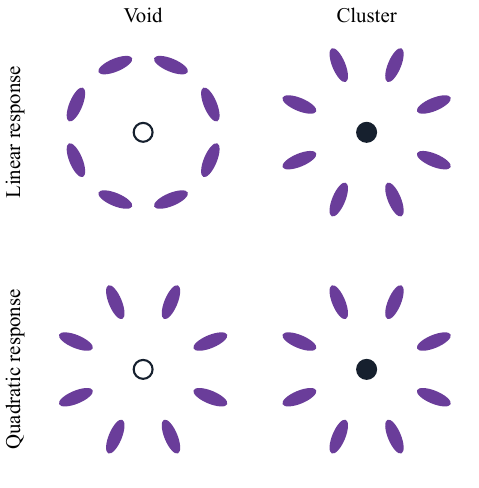}{The tidal field around a void and a cluster}
  \caption{\textbf{The alignment patterns from linear and quadratic alignment models.} Shapes of galaxies (purple) around a void (open circle) and a cluster (filled circle). The top row shows the linear response to the tidal field where the shape alignment switches from tangential to radial between voids and clusters and the bottom row the quadratic response where the signal must keep the same sign for both voids and clusters.}
  \label{fig:tidal}
\end{figure}

$C_1$ is often referred to as the tidal alignment model, and terms proportional to $C_2$ are often referred to as the tidal torquing model. The term $C_{1\delta}$ only appears when considering the inhomogeneous sampling of the tidal field from galaxies. Together, these terms form the tidal alignment and tidal torquing (TATT) model. Physically, however, this is a misnomer: torquing models can also produce a linear response to the tidal field\cite{HuiZhang2008,LarsenChallinor2016,Blazek2019}. Once the density field has evolved non-linearly, a large-scale mode modulates the growth of the small-scale tidal field that torques the galaxy, and the resulting alignment contains a term that responds linearly to the large-scale tidal field. By classic tidal torque theory we mean the case without this effect, in which the alignment is purely quadratic in the initial, Gaussian tidal field (\mref{sec:model}). Therefore we refer to $C_1$ as the linear response to the tidal field, $C_{1\delta}$ as the density-weighting response, and $C_2$ as the quadratic response. The galaxy ellipticity is defined as $e=|e|e^{2i \psi} = e_1+i e_2$, where the factor of two comes from the fact that a rotation of an ellipse by $\pi$ gives an identical ellipse. The galaxy ellipticity is a spin-2 quantity. The possible alignment patterns of galaxies are radial~\iarad\ ($e_+>0$), tangential~\iatan\ ($e_+<0$) and cross~\iacross\ ($e_\times\neq0$). In Fig.~\ref{fig:tidal} we illustrate the alignment of galaxy shapes from linear and quadratic response. The linear response switches sign between a void and a cluster, whereas the quadratic response does not.

Galaxy alignments (or intrinsic alignments) are a crucial systematic for weak gravitational lensing surveys\cite{HirataSeljak2004,Joachimi2015,Troxel2015}. Weak gravitational lensing describes the deformation of galaxy images on the sky through the deflection of light by massive objects. These deformations lead to correlations between galaxy shapes, and between galaxy shapes and positions. Correlations from intrinsic alignments can therefore be mistaken for correlations from lensing. It is for this reason that it is essential to improve our understanding of intrinsic alignments of galaxies for precise weak gravitational lensing cosmology.  

There have been many efforts to measure galaxy alignments for different galaxy types. For red galaxies, significant radial alignments of galaxies have been found consistently\cite{Mandelbaum2006,Hirata2007,Joachimi2011,Singh2015,Johnston:2018nfi,Yao:2020jpj,Georgiou:2025pca,DESI:2025fda}. These observations have been well described by the linear response model. For blue galaxies and late-type galaxies~\iaspiral, the observational situation is more complex. Within the field of galaxy formation there have been detections of blue galaxy spin alignments with filaments of the cosmic web \cite{Tempel2013,Welker2020}, however, using integral-field kinematics, other authors found inconsistent results \cite{Krolewski2019}. These results use the reconstructed cosmic web, instead of the alignment of galaxies around other galaxies, therefore their importance for weak gravitational lensing is difficult to assess. With an approach more relevant to weak lensing surveys, blue galaxies have given null detections\cite{Mandelbaum2011,Johnston:2018nfi,Yao:2020jpj,Georgiou:2025pca,DESI:2025fda}, until the recent $3.3\sigma$ hint of tangential alignment for massive blue galaxies in \refcite{Paviot2026}.

While the linear alignment model for early-type galaxies is well established, both in theory and in observations, to produce radial alignments around overdensities, the predictions for disc galaxies are far less clear. Many hydrodynamical simulations studies find that disc or blue galaxies have either no detectable large-scale alignment\cite{Tenneti2016,Hilbert2017} or a radial allignment\cite{Samuroff2021,Zjupa2022,Delgado2023,vanHeukelum2025}. In the Horizon-AGN simulation, however, disc galaxies have large-scale tangential alignments in 3D\cite{Chisari2015}, but the projected alignment of theses discs is consistent with zero. Furthermore the alignment of simulated galaxies within the cosmic web appears to depend on the particular hydrodynamical code used for the simulation \cite{Tenneti2016}. These simulations are currently limited by the simulation volume and the difficulty of creating realistic disc galaxies within large volume simulations \cite{Chisari2015,Chisari2016}.

One reason for the strong detections for luminous red galaxies (LRGs) and the lack of detections for blue galaxies is the selection inherent to large spectroscopic galaxy redshift surveys, which have generally targeted red early-type galaxies~\iaelliptical, such as LRGs, which reside in massive halos. These galaxies are excellent tracers of the large-scale structure for cosmology, as they are bright, allowing deeper selections in redshift, and strongly clustered, giving a large signal in their galaxy auto-correlation functions. This has biased the literature of direct alignment measurements away from the blue discs which we study in this work. Due to the relatively weak constraints on the alignments of blue galaxies, they have been assumed to have small or no intrinsic alignment signals\cite{Joachimi2011}. The majority of galaxies used as weak gravitational lensing sources for the upcoming Euclid\cite{Mellier2025}, LSST\cite{Ivezic2019} and Roman\cite{Spergel2015} surveys will be blue star-forming galaxies. Therefore it is crucial that we understand the amplitudes of their alignments, so that cosmological inferences are not biased.

In order to measure galaxy alignments, we measure the cross-correlation between galaxy shapes and galaxy positions and the auto-correlation of galaxy positions. The cross-correlation between galaxy shapes and positions, $w_{g+}(r_p)$, is closely related to the mean shape of galaxies, $\langle e_+\rangle$, at projected separation $r_p$ from another galaxy, integrated along the line of sight.

\begin{equation}
  w_{g+}(r_p) = b_g\big[A_1\,L(r_p) + A_{1\delta}\,D_{\rm w}(r_p) + A_2\,Q(r_p)\big] ,
\label{eq:wgp}
\end{equation}

where $A_1$, $A_{1\delta}$ and $A_2$ are dimensionless versions of $C_1$, $C_{1\delta}$ and $C_2$, which make it possible to compare the amplitudes across different redshifts (the relations between $A$ and $C$ are given in \mref{sec:model}, Eq.~\eqref{eq:C1C2}). Positive $A_1$ leads to radial alignments and negative $A_1$ leads to tangential alignments. $L$, $D_{\rm w}$ and $Q$ are functions that depend upon the assumed cosmology and the redshift distributions of the shape and position samples; they are defined in \mref{sec:model} (Eqs.~\eqref{eq:calA}--\eqref{eq:calB}).

The auto-correlation of galaxy positions measures the galaxy bias,

\begin{equation}
  w_{gg}(r_p) = b_g^2\, G(r_p) ,
\label{eq:wgg}
\end{equation}

where $G$ is the projected matter correlation function; the full expression is given in \mref{sec:model} (Eq.~\eqref{eq:wggproj}). By combining measurements of $w_{gg}$ and $w_{g+}$ at fixed cosmology we can directly constrain the amplitude of intrinsic alignment. Unless stated otherwise, we fit our measurements with the linear response alone, setting $A_{1\delta}=A_2=0$.

\subsection*{Blue discs show tangential alignments}

\begin{figure}
  \centering
  \placefig{\columnwidth}{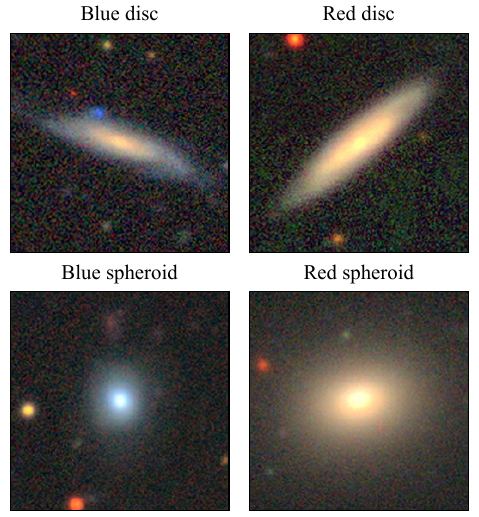}{Example galaxies of each sample}
  \caption{\textbf{An example galaxy from each of the shape samples.} Galaxy images using $grz$ bands from the Legacy surveys. Each image is $38''$ on a side, and all galaxies are within the redshift range $0.05<z<0.08$.}
  \label{fig:galaxies}
\end{figure}

\begin{figure}[htbp]
  \centering
  \placefig{\columnwidth}{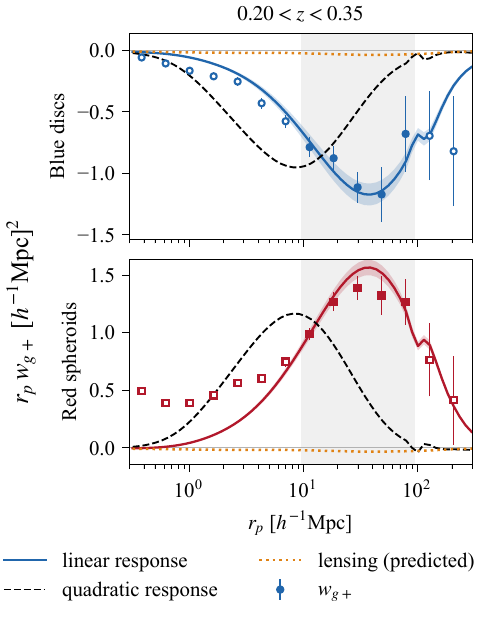}{Blue discs and red spheroids at $0.20<z<0.35$ with LA and TT}
  \caption{
  \textbf{Blue discs align tangentially and red spheroids align radially.} The cross-correlation of galaxy shapes with galaxy positions, $r_p w_{g+}$, for the redshift bin $0.20<z<0.35$. Both samples are well fitted by the linear response (solid line) and poorly fitted by the quadratic response (dashed line). The predicted gravitational lensing signal (dotted line) is very small compared to the observed correlations.}
  \label{fig:shape}
\end{figure}

\begin{figure*}[tbp]
  \centering
  \placefig{\textwidth}{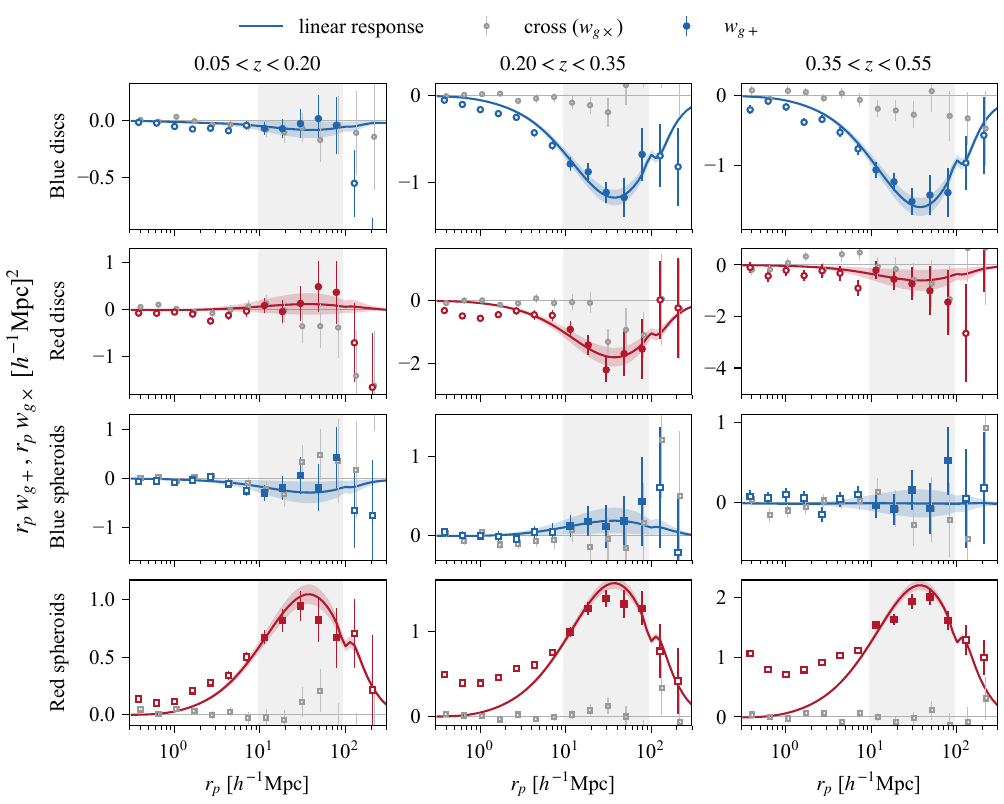}{$w_{g+}$ for each galaxy type, with the LA predictions}
  \caption{\textbf{Alignment signal for each galaxy type in redshift bins.}
  Each of the alignment signals except for the red spheroids at $0.35<z<0.55$ are well modelled by the linear response. The cross component is consistent with a null detection throughout, as expected for both intrinsic alignment and weak lensing measurements at this precision.
  $r_p w_{g+}$ (coloured points) and the cross component $r_p w_{g\times}$ (grey open points) of blue discs, red discs, blue spheroids and red spheroids (rows) in the three redshift bins (columns). The grey band shows the range of scales used for the model fit, $10\le r_p\le100\,h^{-1}$\,Mpc. Line and band: the linear-response fit and its 68\% interval, including the predicted lensing. The errors are estimated using a 100-region jackknife.
  }
  \label{fig:wgp}
\end{figure*}
\begin{figure*}[tbp]
  \centering
  \placefig{\textwidth}{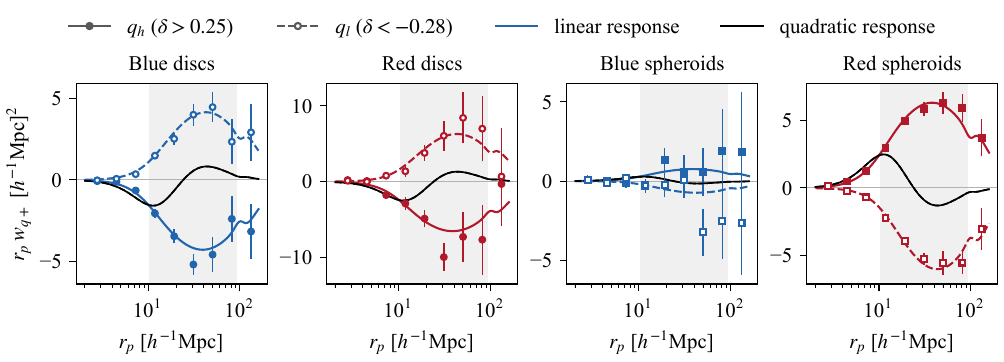}{$w_{q+}$ around overdensities and voids, with the linear and quadratic responses}
  \caption{\textbf{The alignment switches sign between voids and clusters, which cannot be explained by the quadratic response.}
 $r_p w_{q+}$ around random points in the densest 20\% of the density field smoothed on $5\,h^{-1}$\,Mpc, at $0.20<z<0.35$ ($q_h$, $\delta>0.25$; filled points,
  solid lines) and around the emptiest 20\% of the field ($q_l$, $\delta<-0.28$; open points, dashed lines). The coloured lines show the best fitting linear response. Black lines show the quadratic response with its amplitude set by the densest 20\% at $10\,h^{-1}$\,Mpc. The models are described in \mref{sec:environment}. The quadratic response must have the same alignment pattern around underdense and overdense regions, which is in clear contradiction with our blue disc and red spheroid observations.}
  \label{fig:environment}
\end{figure*}
\begin{figure}[htbp]
  \centering
  \placefig{\columnwidth}{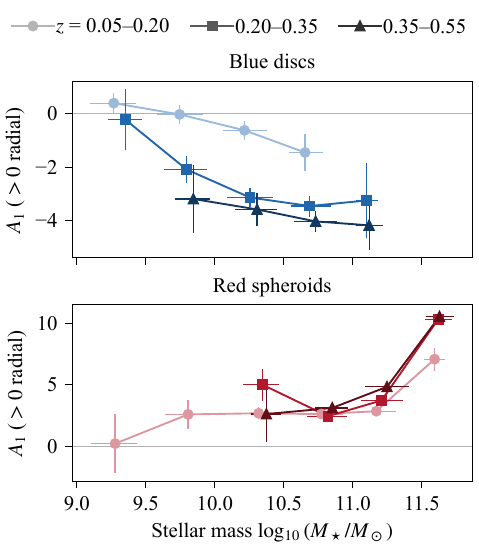}{Blue discs and red galaxies as a function of stellar mass}
  \caption{\textbf{The alignment signal strengthens with stellar mass.} Here we show the amplitude of the linear response $A_1$ for samples in bins of stellar mass.
  \textbf{a}, blue discs and \textbf{b}, red spheroids. The marker and the shade of the colour show the redshift bin with the darker shades showing higher redshift. Horizontal bars span 16--84\% of each cell's masses. }
  \label{fig:mass}
\end{figure}
 
In order to study the alignments of galaxies within the cosmic web, we require precise distances to each of the galaxies. We selected resolved sources with an $r$-band magnitude $r<20.5$ from the DESI Legacy Imaging Surveys and obtained their spectrophotometry using data from the SPHEREx satellite\cite{Dore2014,Bock2025}. From this we obtained a redshift catalogue of 71.2 million galaxies with a $\sigma_{\rm NMAD}$ uncertainty of 0.0065 and 0.46\% catastrophic outliers across the full sky (\mref{sec:catalogue}). This sample has no colour or morphological selection, so we can measure intrinsic alignment signals for any type of galaxy. A full-sky projection of the cosmic web from this catalogue is shown in Fig.~\ref{fig:cosmicweb}. From this redshift catalogue, we selected a high-quality redshift subsample by keeping the 90\% of galaxies with the narrowest redshift probability distributions. This leaves us with a sample of 63.0 million galaxies, with $\sigma_{\rm NMAD}=0.0055$--0.0077 and 0.13--0.52\% catastrophic outliers in the three redshift bins of our analysis.

We divide this sample into disc-type galaxies and spheroidal galaxies, using the S\'ersic index measurements from the DESI Legacy Imaging Surveys. These two groups are further subdivided into a red sample, composed of dusty galaxies filled with old stars, and a blue sample, composed of star-forming galaxies filled with young stars, using a colour cut in thin redshift slices (\mref{sec:samples}; Extended Data Fig.~\ref{fig:colour}). For our shape samples we use shape measurements from the Tractor; we repeat our measurements with DES Y6 shapes in Extended Data Fig.~\ref{fig:desshapes} (\mref{sec:des}), where we show that the two shape measurement methods give consistent results. Example galaxies of each sample are shown in Fig.~\ref{fig:galaxies}. Table~\ref{tab:samples} shows the number of galaxies in each of these shape samples for each redshift bin; the position samples contain 11.0, 16.1 and 12.1 million galaxies in the three bins. We then measured the cross-correlation of each of these galaxy shape samples with the full galaxy position sample.

In Fig.~\ref{fig:shape} we see clear tangential alignment for blue disc galaxies and the opposite, radial alignment for red spheroidal galaxies. Both samples are well described by a linear response to the tidal field (probabilities to exceed, PTE, of 0.58 and 0.59) and poorly described by a quadratic response alone (PTE of $3\times10^{-6}$ and $7\times10^{-25}$). The quadratic response model has the wrong shape: it peaks near $10\,h^{-1}$\,Mpc and then decreases too rapidly to explain the shape of the measured correlation function. For each model we fit between 10 and 100\,$h^{-1}$\,Mpc. The cut at $10\,h^{-1}$\,Mpc avoids the regime where structure formation becomes non-linear. We fit $w_{g+}$ and $w_{gg}$ at the same time. The models are described in \mref{sec:model}, the galaxy auto-correlation measurements in \mref{sec:clustering} (Extended Data Fig.~\ref{fig:wgg}), and the fitting in \mref{sec:fitting}. The predicted gravitational lensing signal, given the stellar masses of these galaxies and their redshift distribution, is only 3\% of the observed blue-disc correlation function. Therefore lensing cannot explain this observation. The inconsistency of lensing with this observation is further explored in \mref{sec:lensing}.

In Fig.~\ref{fig:wgp} we show the alignment correlation functions for each colour, morphological type and redshift bin. We use two measures of the detection significance. The first is the $\chi^2$ of the measured $w_{g+}$ against zero over all scales, $0.4<r_p<335\,h^{-1}$\,Mpc, which tests the consistency of the observation with no alignment and does not depend on a model. The second is the amplitude of the linear response divided by its uncertainty, $|A_1|/\sigma_A$ (Table~\ref{tab:fits})\footnote{Above $\sim\!10\sigma$ the detection significance will depend on the tails of the estimated covariance. As we do not expect this to be extremely precise at this level, detections above $10\sigma$ should be taken as indicative rather than exact.}. For the intermediate and high redshift bins, blue discs~\iaspiral\ show strong tangential alignment signals~\iatan[iablue]. For the intermediate redshift bin this is inconsistent with no alignment at $15.7\sigma$, and for the high redshift bin at $16.2\sigma$ ($22.5\sigma$ combined; $|A_1|/\sigma_A=12.4$ and 11.6). In the lowest redshift bin blue disc galaxies show a much smaller alignment signal; as shown in Fig.~\ref{fig:mass}, this is possibly because the lowest redshift bin contains the blue disc sample with the lowest mass on average. The intermediate redshift bin for the red discs shows an $8.4\sigma$ detection of tangential alignment, the same sign as the blue discs. However, no signal is seen in the other two redshift bins. Red spheroidal galaxies~\iaelliptical\ show strong radial alignments~\iarad[iared] in every redshift bin, at $16.2\sigma$, $42.4\sigma$ and $46.8\sigma$ ($65.0\sigma$ combined; $|A_1|/\sigma_A=12.2$, 22.9 and 27.1), while blue spheroids show no alignment in any bin.

The cross component $w_{g\times}$~\iacross, which is expected to be very small for intrinsic alignments, is consistent with zero for every measurement. A non-zero $\times$ component would indicate systematic issues with the measurement, therefore this is a reassuring quality check. There are many systematic issues that would produce only a $+$ signal; these are tested in Methods (\nameref{sec:corr}, \nameref{sec:lensing} and \nameref{sec:des}), where we find no evidence of significant systematic effects that can change the conclusions of this work.

The linear response alone provides a good fit to every measurement except the red spheroids at $0.35<z<0.55$, with probabilities to exceed between 0.57 and 0.99 elsewhere. We discuss the red spheroids at $0.35<z<0.55$ in the Discussion and in \mref{sec:amplitudes}. The quadratic response alone is rejected for every measurement with a significant detection. For the blue spheroids we do not detect an alignment signal for any of the redshift bins. The red spheroids in our highest redshift bin are poorly fit by the linear model (PTE 0.001), but they are fitted well when the linear matter power spectrum is replaced with the non-linear matter power spectrum (the Non-linear Linear Alignment (NLA) model \cite{bridle2007}; PTE 0.12). Therefore we expect that the linear model that we use breaks down, within this redshift bin, which is the redshift bin containing the most massive galaxies within our sample (given the flux-limit of the galaxy sample).

The tangential alignment of blue discs cannot be explained by gravitational lensing as discussed in \mref{sec:lensing}. 

\subsection*{The alignment switches from tangential to radial between clusters and voids}

We can provide further evidence that the response is linear in the tidal field by considering how galaxies align around overdense and underdense regions. A quadratic response must produce alignments of the same sign around overdense and underdense regions, whereas for a linear response tangential alignments around clusters would imply radial alignments around voids. We test this directly with our observations. We estimate the galaxy density field, $\delta_g = n_g/\bar{n} -1$, from all galaxies of the position sample at $0.20<z<0.35$, smoothed with a Gaussian of $5\,h^{-1}$\,Mpc across the line of sight and $30\,h^{-1}$\,Mpc along the line of sight. We then place 2 million random positions within the field, and select those within the top 20\% of density values, $q_h$, and those within the lowest 20\%, $q_l$. We then measure the alignment signals $w_{q_h+}$ and $w_{q_l+}$ around each of these position samples in the same manner as for galaxy positions (\mref{sec:environment}). These correlation functions are presented in Fig.~\ref{fig:environment}.

The sign of the alignment signal switches between underdensities and overdensities for each of the samples with a detected alignment signal. Red spheroids align tangentially around voids~\iatanv[iared] and radially around clusters~\iarad[iared]. Blue discs align radially around voids~\iaradv[iablue] and tangentially around clusters~\iatan[iablue]. It is not possible to explain these observations with a quadratic response or with density weighting; therefore we can rule out that the alignments of disc galaxies are dominated by a physical mechanism whose alignment keeps the same sign when the density field changes sign. Fitting both regions together, the linear response describes the blue discs with $A_1=-2.43\pm0.22$ (PTE 0.80), whereas the quadratic response alone does not (PTE $7\times10^{-23}$). The model used here is explained in \mref{sec:environment}.

\subsection*{The alignment is stronger at higher stellar mass}

We can now extend the study further to look at the dependence of the alignment amplitudes on galaxy stellar mass. In order to estimate stellar masses we ran CIGALE\cite{Boquien2019} on the LSx catalogue, using the DESI Legacy Imaging photometric bands and the SPHEREx spectrophotometry (\mref{sec:mass}).

In Fig.~\ref{fig:mass} we see that both the blue disc and red spheroid alignments are strongly correlated with galaxy stellar mass. The red spheroids show a plateau at low stellar masses and a steep rise at the high stellar mass end, in agreement with multiple other studies\cite{Hirata2007,Joachimi2011,Singh2015,HervasPeters2025,Paviot2026}. The blue disc tangential alignment increases significantly at high stellar mass. The blue disc sample does not reach the high stellar masses of the red spheroid sample. We find that for both discs and spheroids the most massive galaxies have the strongest alignment signals.

\subsection*{Discussion}

The significant detection of the tangential alignments of blue disc galaxies that we observe in this work does not have an adequate explanation from theoretical models or hydrodynamical simulations. The amplitude of alignment is large, $A_1=-2.81\pm0.23$ and $-3.79\pm0.33$ at $0.20<z<0.35$ and $0.35<z<0.55$, and comparable in size to the amplitude of radial alignments of red spheroids. The measurements cannot be explained by redshift outliers, gravitational lensing contamination or the shape measurement method (Methods: \nameref{sec:lensing} and \nameref{sec:des}). The blue alignment signal is well described by a linear response to the tidal field. The alignment switches sign between underdensities and overdensities, which provides further evidence that the mechanism creating these alignments produces a linear response to the tidal field, and not a quadratic response, which has previously been the theoretically preferred model for the alignment of disc galaxies\cite{Catelan2001,LeePen2000,HirataSeljak2004}.

Past measurements of the alignments of blue galaxies found no alignment\cite{Mandelbaum2011,Johnston:2018nfi,Yao:2020jpj,Georgiou:2025pca,DESI:2025fda}, until \refcite{Paviot2026} measured a $3.3\sigma$ hint of tangential alignment for massive blue galaxies. There are two reasons for this. First, the large spectroscopic surveys used for direct alignment measurements target red galaxies, so samples of blue galaxies with precise redshifts have been small. Second, blue samples have been selected by colour alone, without a morphological selection, which dilutes the alignment signal of the discs with that of other blue galaxies. As the LSx catalogue has no colour selection we are able to probe the alignment of any type of galaxy, provided that they are bright enough to enter the sample. We see this dilution in Extended Data Fig.~\ref{fig:blueall}: when the shape sample is selected on galaxy colour alone, the alignment amplitude is significantly smaller. For the redshift bin $0.20<z<0.35$ all resolved blue galaxies have an alignment amplitude of $A_1=-2.06\pm0.19$, compared with $-2.81\pm0.23$ for the blue disc galaxies. Therefore it is specifically blue disc galaxies which show the strongest alignment signals. Our blue galaxy sample is large enough, and our redshifts are precise enough that even when measuring the ensemble of blue galaxies we have a significant detection of tangential alignments. 

The detection of the linear tangential alignments of disc galaxies contradicts classic tidal torque theory\cite{Catelan2001,LeePen2000,HirataSeljak2004}, in which the alignment is quadratic in the Gaussian tidal field and its cross-correlation with galaxy positions should only give a weak quadratic alignment\cite{HirataSeljak2004}. This is also in contradiction with hydrodynamical simulations, in which disc galaxies show either no large-scale alignment or radial alignments\cite{Tenneti2016,Hilbert2017,Samuroff2021,Zjupa2022,Delgado2023,vanHeukelum2025} with the exception of the Horizon-AGN simulations \cite{Chisari2015,Chisari2016}. 

Significant spin alignments of blue discs with their host filament have been found in both observations\cite{Tempel2013,Welker2020} and simulations\cite{Codis2015,Codis2018,Kraljic2020}. Low-mass disc galaxies have spins aligned to their filament, and more massive galaxies have perpendicular spins\cite{Codis2018,Welker2020}. The disc major axis is perpendicular to the disc spin. As filaments are connected to clusters, this would naively lead to tangential alignments for low-mass discs and radial alignments for high-mass discs, the converse of what we observe (Fig.~\ref{fig:mass}). 
 
Therefore the physical mechanism responsible for such strong tangential alignments in disc galaxies is unclear. The measurements do not exclude more complicated tidal-torquing mechanisms. For example, non-linear evolution can produce an effective linear response from tidal-torquing \cite{HuiZhang2008,Blazek2019}. On large-scales this cannot be distinguished from the linear response that we measure. Therefore torquing may be the cause of the linear alignment that we observe\cite{HuiZhang2008,LarsenChallinor2016,Blazek2019}. \refcite{Blazek2019} does not make predictions for the sign of this physical mechanism and \refcite{LarsenChallinor2016} found that it should be radial, in contradiction with our observations. Furthermore this effect has not be found in hydrodynamical simulations, so this remains a tentative explanation. More theoretical research is needed to understand the sign, amplitude and mass dependence of these results.

These results are important for future weak gravitational lensing measurements. The majority of source galaxies for weak lensing in Euclid\cite{Mellier2025}, LSST\cite{Ivezic2019} and Roman\cite{Spergel2015} will be blue star-forming galaxies. We have shown that large blue disc galaxies align with an amplitude comparable to that of red spheroids. In Extended Data Fig.~\ref{fig:desshapes} (\mref{sec:des}) we also see that small blue galaxies align radially. Therefore the intrinsic alignment signal within a lensing survey will depend sensitively upon the source selection of galaxies. It is therefore essential that intrinsic alignment models are calibrated for the specific sample used for the lensing sources, in terms of survey depth, galaxy size and galaxy morphology. For surveys as deep as Euclid\cite{Mellier2025}, LSST\cite{Ivezic2019} and Roman\cite{Spergel2015} this will be observationally challenging. Alternatively we need to develop a much deeper physical understanding of how disc galaxies acquire the large tangential alignments that we have found in this work.

\FloatBarrier
\renewcommand{\figurename}{Extended Data Fig.}
\setcounter{figure}{0}
\renewcommand{\theHfigure}{ED\arabic{figure}}%

\section*{Methods}

\subsection*{The Legacy Surveys--SPHEREx redshift catalogue}\label{sec:catalogue}

To create a high quality redshift catalogue (Legacy x SPHEREx, LSx catalogue) we select sources using the DESI Legacy Imaging Surveys and perform forced photometry in SPHEREx images for each source. The DESI Legacy Imaging Surveys\cite{Dey2019} (DR9 north, DR10 south and DR11 south) combine imaging from several telescopes, processed in a consistent manner using the Tractor\cite{Lang2016}. As noted in the main text, our selection is deliberately simple: we select sources with dereddened $r<20.5$, at least one observation in each of the $g$, $r$ and $z$ bands, and we exclude sources that are identified by the Tractor as unresolved (PSF morphological type). This last cut primarily removes stars, and more than halves the number of sources to be measured. SPHEREx\cite{Bock2025} is a near-infrared satellite that surveys the entire sky in 102 bands between 0.75--5.0\,$\mu$m. By combining the Legacy imaging source selection with SPHEREx we can measure precise redshifts. For this catalogue we used SPHEREx observations between 24 April 2025 and 20 July 2026. 

The LSx catalogue was validated using spectroscopic redshifts from DESI DR1\cite{DESI2026DR1}, SDSS\cite{Abdurrouf2022}, GAMA\cite{Driver2022}, 6dFGS\cite{Jones2004,Jones2009} and 2dFGRS\cite{Colless2001}. Before redshift quality cuts the sample has a redshift uncertainty of $\sigma_{\rm NMAD}=0.0065$ and an outlier rate of 0.46\%, where outliers are defined as galaxies with $|\Delta z|>0.15$ and $\Delta z=(z_{\rm est}-z_{\rm spec})/(1+z_{\rm spec})$. 
 
\subsection*{Position and shape samples}\label{sec:samples}

\paragraph{Position sample.} To define a high-quality position sample for galaxy clustering, we first apply the redshift quality cut, keeping the 90\% of galaxies with the narrowest redshift probability distributions (63.0 million galaxies). We then remove regions of high Galactic extinction ($E(B-V)>0.10$\cite{Chiang2023}), high stellar density (above 2000\,deg$^{-2}$ for bright Gaia sources $G<16.5$), and regions within $8^\circ$ and $5^\circ$ of the Large and Small Magellanic Clouds. We do not include the DR11 south region of the Legacy Imaging Surveys as we found higher contamination of the catalogue within this area. After removing these regions, the clustering sample contains 39.2 million galaxies with $0.05\le z_{\rm est}<0.55$, over 19,312\,deg$^2$. We create systematic weights for each galaxy using the iterative systematics decontamination method\cite{RodriguezMonroy2022}. This removes the linear dependence of the galaxy density on $E(B-V)$, stellar density, the $g$, $r$ and $z$ depths and the $r$-band seeing. The probability that a given galaxy passes the quality cut of the clustering catalogue selection depends upon the SPHEREx depth, which varies significantly across the sky, due to increased photon noise in regions of higher zodiacal background light. To correct for this we also weight the randoms with a linear correction based upon the SPHEREx depth which removes this dependence. 

\begin{figure*}[tbp]
  \centering
  \placefig{\textwidth}{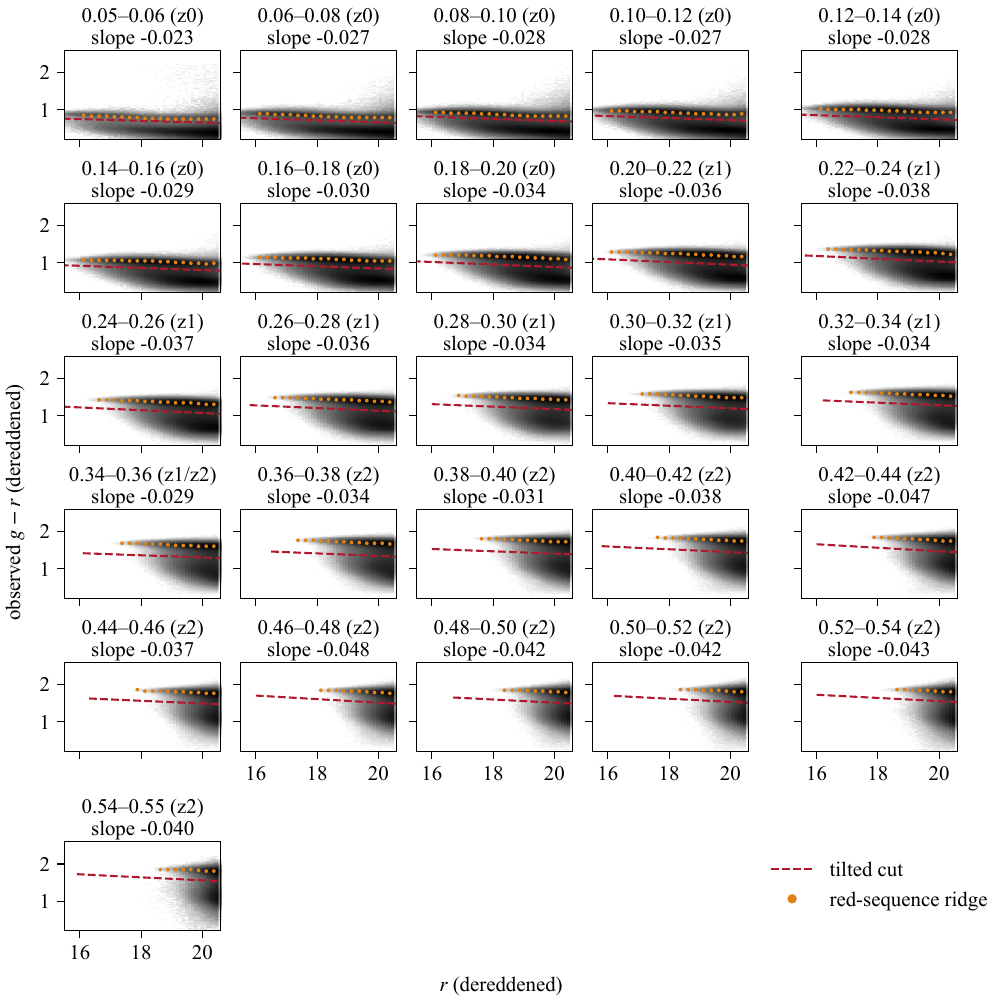}{Colour--magnitude diagrams of every redshift slice with the tilted cut}
  \caption{\textbf{The red-blue galaxy separation is clear for LSx. }
  The dereddened $g-r$ colour against $r$ magnitude for every galaxy in each $\Delta z=0.02$
  slice of redshift (grey, logarithmic counts). The red sequence is shown with the dotted orange line. The cut we use between the blue and red galaxies is parallel to the red-sequence.  }
  \label{fig:colour}
\end{figure*}

\paragraph{Shape samples.} For the shape sample we select galaxies identified as Tractor SER, EXP or DEV sources, with a finite ellipticity $|e|<1$ and a resolved shape, ${\rm res}\equiv r_{\rm half}/{\rm FWHM}_{\rm PSF}>1$, where $r_{\rm half}$ is the half-light radius of the Tractor model (SHAPE\_R) and ${\rm FWHM}_{\rm PSF}$ is the full width at half maximum of the $r$-band point-spread function at the position of the source (PSFSIZE\_R). This leaves us with 5.61, 4.78 and 3.21 million galaxies in each redshift bin. We select a disc sample, as galaxies with either EXP morphological type or SER models with S\'ersic index $n<1.5$. We select a spheroid sample, as galaxies with either DEV morphological type or SER models with S\'ersic index $n\geq2.5$. SER models with $1.5\leq n<2.5$, 15, 14 and 12\% of the shape sample in the three redshift bins, are in neither sample, so the four samples of Table~\ref{tab:samples} contain 4.78, 4.10 and 2.84 million galaxies. Each sample is then split into red and blue using the dereddened $g-r$ colour (Extended Data Fig.~\ref{fig:colour}). 

The number of galaxies in each sample is given in Table~\ref{tab:samples}. We show the $n(z)$ for each of the shape samples in Extended Data Fig.~\ref{fig:photoz}. 

\paragraph{Ellipticities.} We use the Tractor ellipticities SHAPE\_E1 and SHAPE\_E2. The Tractor forward models each source with different light profiles; a point source, a round exponential, an exponential, a de Vaucouleurs or a S\'ersic profile, each convolved with the point-spread function. If a source is well-fit by a round exponential then no shape is measured, therefore these sources are excluded from our catalogue. With this forward-modelling method the shapes are corrected for the point-spread function, but they are not calibrated for weak lensing shear. We show in \nameref{sec:des} that our results are robust against the shape measurement method used. An additional issue is that the Tractor chooses the round exponential when an ellipticity does not lead to a significantly better fit, which is primarily the case for small and faint sources. These objects make up 5--7\% of our shape samples. Their exclusion means that our selection has a small dependence upon $|e|$. This could lead to a small change in the measured amplitude, but it will not affect the detection. For example, with DES Y6 shapes, which do not exclude round galaxies, we still obtain significant tangential alignments for blue galaxies (Extended Data Fig.~\ref{fig:desshapes}).

\subsection*{Stellar masses}\label{sec:mass}

We estimate stellar masses for the LSx catalogue using CIGALE\cite{Boquien2019}. CIGALE estimates the galaxy stellar mass using spectral energy distribution models based upon the galaxy's star-formation history, stellar populations, nebular emission, dust attenuation and dust emission. We use the same models that were used for the CIGALE value-added catalogue of the DESI Early Data Release\cite{Siudek2024}. The stellar mass fit succeeds for 97\% of the quality sample. Compared with the DESI DR1 CIGALE catalogue\cite{DESI2026DR1}, our stellar masses have a median offset of $+0.04$\,dex and a scatter of $\sigma_{\rm NMAD}=0.16$\,dex.

\subsection*{Correlation functions}\label{sec:corr}

In this work we measure galaxy position-shape correlations and galaxy position--position correlations. For the position-shape correlations we use the same estimator as \refcite{Singh2015},

\begin{equation}
  \xi_{g+}(r_p,\Pi) = \frac{S_+D - S_+R}{SR} ,
\label{eq:estimator}
\end{equation}

where $S_+D$ is the sum of $e_+$ over shape--position pairs at projected separation $r_p$ and line-of-sight separation $\Pi$, $S_+R$ is the same sum around random positions, and $SR$ is the shape random counts. The subtraction of $S_+R$ removes potential additive shape measurement systematics; we find that our $S_+R$ measurements are consistent with zero on the scales that we fit.

For the galaxy autocorrelation, $w_{gg}$, we use the Landy--Szalay estimator\cite{LandySzalay1993}

\begin{equation}
  \xi_{gg}(r_p,\Pi) = \frac{DD - 2DR + RR}{RR} ,
\label{eq:ls-estimator}
\end{equation}

where $DD$, $DR$ and $RR$ are the position--position, position--random and random--random pair counts at a separation $(r_p,\Pi)$, each normalised by the weighted total number of pairs. We then integrate each of the correlation functions over the line-of-sight between $-\Pi_{\max}$ and $\Pi_{\max}$ with  $\Pi_{\max} = 100\,h^{-1}$\,Mpc, to measure the projected correlation function $w_{g+}(r_p)$ and $w_{gg}(r_p)$. 

We measure each correlation function with 15 logarithmic bins in physical separation $r_p$, from 0.3 to 427\,$h^{-1}$\,Mpc. For each of the correlation functions we use TreeCorr\cite{Jarvis2004}. The position sample is weighted using the systematic weights described in \nameref{sec:samples}. The shape sample is unweighted. Using inverse-variance weights for the shape sample changes the inferred amplitude $A_1$ of the blue discs at $0.20<z<0.35$ by $6\times10^{-4}$. This is because the measurement uncertainty, $\sim0.004$, is much smaller than the intrinsic scatter, $\approx0.30$.

For each correlation function measurement we estimate the covariance using a jackknife with 100 angular regions. The same angular regions are used for every correlation function, so that we can correctly calculate the covariance between different measurements. We correct the inverse covariances using the Hartlap factor\cite{Hartlap2007}.

\subsection*{Modelling linear and quadratic responses to the tidal field}\label{sec:model}

In this work we use the tidal alignment and tidal torquing (TATT) model\cite{Blazek2019} to model our observations. As noted in the main body of the text, this model has a somewhat confusing name, but correctly models the physics that we are interested in.

In Eq.~\eqref{eq:response}, we sketched out each of the terms of our model. Following \refcite{Blazek2019}, the full expression is,

\begin{equation}
  \gamma_{ij}=C_1s_{ij}+C_{1\delta}\,\delta\,s_{ij}+C_2\Big(s_{ik}s_{kj}-\tfrac13\delta_{ij}\,s_{kl}s_{kl}\Big) ,
\label{eq:tatt}
\end{equation}

where $\delta$ is the matter density,  $\delta_{ij}$ the Kronecker delta, repeated indices are summed, and $s_{ij}$ is the tidal field. In Fourier space the tidal field is $s_{ij}(\mathbf k)=(k_ik_j/k^2-\delta_{ij}/3)\,\delta(\mathbf k)$. The first term, $C_1 s_{ij}$, is the linear response of observed galaxy shapes to the tidal field. The second term, $C_{1\delta}\,\delta s_{ij}$, arises because we necessarily sample the shape field at the positions of galaxies, so the tidal field is weighted by the local density. The third term, $C_2\, s_{ik}s_{kj}$, is the quadratic response. Following \refcite{Blazek2019}, we can write the three coefficients in terms of dimensionless amplitudes,

\begin{equation}
  \begin{split}
  C_1 &= -\frac{A_1 \bar C_1 \rho_{\rm crit}\Omega_{\rm m}}{D(z)} , \\
  C_{1\delta} &= -\frac{A_{1\delta} \bar C_1 \rho_{\rm crit}\Omega_{\rm m}}{D(z)} , \\
  C_2 &= \frac{5A_2 \bar C_1 \rho_{\rm crit}\Omega_{\rm m}}{D^2(z)} ,
  \end{split}
\label{eq:C1C2}
\end{equation}

where the parameters that we choose to measure in our analyses are $A_1$, $A_{1\delta}$ and $A_2$. $D(z)$ is the linear growth factor, normalised to unity today, and $\bar C_1\rho_{\rm crit}=0.0134$. Throughout our analysis we use different combinations of $A_1$, $A_{1\delta}$ and $A_2$ as free parameters. If the $A_{1\delta}$ term is generated purely by density weighting then we can set $A_{1\delta}=b_SA_1$, where $b_S$ is the linear galaxy bias for the shape sample. However, like $A_1$ and $A_2$, $A_{1\delta}$ may be induced by other physical mechanisms, so we explore both $A_{1\delta}=b_SA_1$ and $A_{1\delta}$ as a free parameter. 

The $w_{g+}$ correlations that we measure can be modelled by projecting the cross-power spectrum of the matter density and the E-mode of the shape field. Following \refcite{Blazek2019}, we have

\begin{multline}
  P_{\delta E}(k,\mu) = (1-\mu^2)\Big[\overbrace{C_1P_\delta(k)}^{\rm linear} + \overbrace{C_{1\delta}\big(A_{0|0E}+C_{0|0E}\big)(k)}^{\rm density\ weighting} \\
  + \overbrace{C_2\big(A_{0|E2}+B_{0|E2}\big)(k)}^{\rm quadratic}\Big] ,
\label{eq:PdE}
\end{multline}

where for $P_{\delta}(k)$ we use the linear power spectrum, $\mu=k_\parallel/k$ is the cosine of the angle between $\mathbf k$ and the line of sight, and the factor $1-\mu^2$ arises from the projection of the tidal field into the plane of the sky. The full expressions for the density-weighting and quadratic terms can be found directly in \refcite{Blazek2019}. 

For a position sample with linear bias $b_g$, each of the three terms of Eq.~\eqref{eq:PdE} gives a contribution to the three-dimensional correlation between galaxy positions and the $+$ component of galaxy shapes.

\begin{equation}
  \begin{split}
  \xi_{g+}(r_p,\Pi,z) ={}& b_g\,\bar C_1\rho_{\rm crit}\Omega_{\rm m}\Big[A_1\, D(z)\,\xi^{+}_{\rm lin}(r_p,\Pi) \\
  &+ A_{1\delta}\,D^3(z)\,\xi^{+}_{\rm den}(r_p,\Pi) \\
  &- 5A_2\,D^2(z)\,\xi^{+}_{\rm quad}(r_p,\Pi)\Big] ,
  \end{split}
\label{eq:xigp-model}
\end{equation}

where $\xi^{+}_{\rm lin}$, $\xi^{+}_{\rm den}$ and $\xi^{+}_{\rm quad}$ can be calculated from their respective power spectra with\cite{Singh2015},

\begin{align}
  \xi^{+}_{X}(r_p,\Pi) ={}& \int_0^\infty \frac{k_\perp^3\, {\rm d}k_\perp}{2\pi}\, J_2(k_\perp r_p) \nonumber\\
  &\times \int_{-\infty}^{\infty} \frac{{\rm d}k_\parallel}{2\pi}\, \cos(k_\parallel \Pi)\, \frac{P_X(k)}{k^2} ,
\label{eq:xiplus}
\end{align}

$J_2$ is the Bessel function of order two, and $k_\perp^2/k^2=1-\mu^2$ is the projection factor of Eq.~\eqref{eq:PdE}. 

Then we project these to obtain the projected correlation functions which we use to constrain our intrinsic alignment models. Following \refcite{Mandelbaum2011},

\begin{align}
  w_{g+}(r_p) ={}& \int {\rm d}z\, W(z) \int {\rm d}\Pi\; p(\Pi\,|\,z)\, \xi_{g+}(r_p,\Pi,z) \nonumber\\
  &+ w_{gG}(r_p) , \label{eq:wproj}\\
  w_{gg}(r_p) ={}& \int {\rm d}z\, W_{gg}(z) \int {\rm d}\Pi\; p_{gg}(\Pi\,|\,z)\, \xi_{gg}(r_p,\Pi,z) . \label{eq:wggproj}
\end{align}

where $W(z)\propto n_S(z)\,n_D(z)/({\rm d}V_c/{\rm d}z)$, which is normalised to 1, $n_S$ and $n_D$ are the redshift distributions of the shape and position samples and $V_c$ is the comoving volume, $W_{gg}(z)\propto n_D^2(z)/({\rm d}V_c/{\rm d}z)$. $p(\Pi\,|\,z)$ is the probability that a shape--position pair at redshift $z$ and true line-of-sight separation $\Pi$ has an observed separation $|\Pi_{\rm obs}|<\Pi_{\max}$, and $p_{gg}$ is the same for position--position pairs. We estimate the product $W(z)\,p(\Pi\,|\,z)$ directly using Monte Carlo sampling. We draw random shape--position pairs from the two samples, with the position galaxies weighted by their systematic weights, and give each galaxy an independent redshift error $z_{\rm true}-z_{\rm est}$, drawn from spectroscopically matched galaxies of similar estimated redshift, magnitude and colour. We keep the pairs with $|\Pi_{\rm obs}|\le\Pi_{\max}$, weighted by $1/\bar\chi^2$, where $\bar\chi$ is the comoving distance at the observed redshifts. From this we can measure the historgram in bins of true redshift and true line-of-sight separation. As the redshift errors are drawn from the measured distribution, catastrophic outliers and non-Gaussian tails are included within this method. $w_{gG}$ is the galaxy--galaxy lensing term, for which we follow \refcite{Georgiou:2025pca}. From Eq.~\eqref{eq:wproj} we can define the $L$, $D_{\rm w}$ and $Q$ terms from the main text,

\begin{align}
  L(r_p) ={}& \bar C_1\rho_{\rm crit}\Omega_{\rm m} \nonumber\\
  &\times \int {\rm d}z\, W(z)\, D(z) \int {\rm d}\Pi\, p(\Pi\,|\,z)\, \xi^{+}_{\rm lin}(r_p,\Pi) ,
\label{eq:calA}\\
  D_{\rm w}(r_p) ={}& \bar C_1\rho_{\rm crit}\Omega_{\rm m} \nonumber\\
  &\times \int {\rm d}z\, W(z)\, D^3(z) \int {\rm d}\Pi\, p(\Pi\,|\,z)\, \xi^{+}_{\rm den}(r_p,\Pi) ,
\label{eq:calD}\\
  Q(r_p) ={}& -5\,\bar C_1\rho_{\rm crit}\Omega_{\rm m} \nonumber\\
  &\times \int {\rm d}z\, W(z)\, D^2(z) \int {\rm d}\Pi\, p(\Pi\,|\,z)\, \xi^{+}_{\rm quad}(r_p,\Pi) .
\label{eq:calB}
\end{align}

The linear power spectrum is computed with CAMB\cite{Lewis2000} for the Planck 2018 cosmology\cite{Planck2018}. We calculated $A_{0|0E}$, $C_{0|0E}$, $A_{0|E2}$ and $B_{0|E2}$ using FAST-PT\cite{Fang2017}. 

\subsection*{Galaxy clustering correlation measurements}\label{sec:clustering}

As explained in the main body of the text, we must measure the galaxy clustering bias in combination with the $w_{g+}$ correlation functions in order to constrain the amplitudes for intrinsic alignments. We measure the galaxy bias from the auto-correlation $w_{gg}$, for each of the redshift bins. We also require the galaxy bias for the shape samples for one of the alignment models that we explore (where we fix $A_{1\delta}=b_SA_1$). We measured the galaxy bias for each shape sample by using the cross-correlation between the shape and position samples. These correlation functions are presented in Extended Data Fig.~\ref{fig:wgg}. We see that for the position sample the galaxy bias increases with redshift, $b_g=1.08$, 1.21 and 1.37. This is expected for a flux-limited galaxy sample: at larger redshifts we can only see the most luminous galaxies which in turn will be more massive and have a larger galaxy bias. The red spheroids have much larger galaxy bias values ($1.38\pm0.08$, $1.51\pm0.05$ and $1.82\pm0.04$) compared to the blue discs ($b_S=0.96\pm0.08$, $0.97\pm0.04$ and $1.12\pm0.03$) which again is expected as red spheroids are expected to live in more massive halos than the blue discs. The PTE for all nine galaxy correlation functions are greater than 0.3, indicating that linear theory describes these observations well on our scales of interest. For the two higher-redshift bins the measurements appear to lie consistently above the prediction; this is a result of the strong correlations between the $r_p$ bins. As the PTE are acceptable, our model provides a good fit to the observations. 

\subsection*{Alignment amplitudes for discs and spheroids}\label{sec:amplitudes}

In Table~\ref{tab:fits} we show the fits of six different models to every measurement presented in Fig.~\ref{fig:wgp}, in combination with the galaxy clustering measurements presented in Extended Data Fig.~\ref{fig:wgg}. The linear response model (1) fits every measurement with a PTE between 0.57 and 0.99, except the red spheroids at $0.35<z<0.55$ ($\chi^2=17.9$ for 4 degrees of freedom, PTE 0.001). The quadratic response alone, model (4), is rejected for every significant detection. For red spheroids in the highest redshift bin, we can obtain a reasonable fit with models (5) and (6), however this requires a large $A_2$ which is counterbalanced by a large $A_1$ of the opposite sign. We interpret this as non-linear evolution on small scales instead of evidence for a non-zero $A_2$ term. As mentioned previously, this observation obtains an acceptable fit (PTE 0.12) when we use the NLA model \cite{bridle2007}.

\subsection*{Alignments around underdense and overdense regions}\label{sec:environment}

As described in the main text, we smooth the galaxy density field with a Gaussian of width $\sigma_\perp=5\,h^{-1}$\,Mpc in the plane of the sky and $\sigma_\parallel=30\,h^{-1}$\,Mpc along the line of sight. We then generate 2 million random positions within the survey footprint. We compute the smoothed density contrast $\delta_c$ at the position of each random, and the normalised peak height $\nu=\delta_c/\sigma_c$, by dividing the smoothed density contrast by its standard deviation over all random positions. We select the 20\% of positions with the lowest density, the quintile $q_l$, and the 20\% with the highest, $q_h$, following the approach used in density-split statistics in galaxy clustering\cite{Paillas2023,Paillas2024,CuestaLazaro2024}. We then measure the alignment correlations around the positions of each quintile with the estimator of Eq.~\eqref{eq:estimator}, except with the quintile positions instead of galaxy positions.

To model the signal, we first start from the correlation function definition. The mean alignment of shapes around the quintile positions $q$, at a separation $\mathbf r'-\mathbf r$, is

\begin{equation}
  \xi_{q+}(\mathbf r'-\mathbf r)\equiv{\rm E}\Big[\big(1+b_S\delta(\mathbf r')\big)\,\gamma_+(\mathbf r')\;\Big|\;\nu(\mathbf r)\in q\Big] ,
\label{eq:ds-xi}
\end{equation}

where the factor $1+b_S\delta$ appears because we only measure shapes at positions where galaxies are observed. This equation is then projected along the line of sight to calculate the model predictions shown in Fig.~\ref{fig:environment},

\begin{align}
  w_{q+}(r_p)={}&\int_{-\Pi_{\max}}^{\Pi_{\max}}\!{\rm d}\Pi_{\rm obs}\int_{-\infty}^{\infty}\!{\rm d}\Pi\nonumber\\
  &\times\frac{e^{-(\Pi_{\rm obs}-\Pi)^2/2\sigma_{\Delta z,s}^2}}{\sqrt{2\pi}\,\sigma_{\Delta z,s}}\;\xi_{q+}(r_p,\Pi) ,
\label{eq:ds-w}
\end{align}

where $\Pi$ and $\Pi_{\rm obs}$ are the true and observed line-of-sight separations and $\sigma_{\Delta z,s}$ is the redshift error of the shape sample in comoving distance.

Throughout we take the x-axis to be the projected separation between the quintile position and the galaxy shape position and the z-axis as the line of sight. Therefore $\mathbf r'-\mathbf r=(r_p,0,\Pi)$ and $\gamma_+=\gamma_{xx}-\gamma_{yy}$. Writing out explicitly Eq.~13 of \refcite{Blazek2019} (the radial component of Eq.~\eqref{eq:tatt}), we have

\begin{align}
  \big(1+b_S\delta\big)\gamma_+={}&C_1s_++b_SC_1\,\delta\,s_+\nonumber\\
  &+C_2\big(-s_{zz}s_++s_{xz}^2-s_{yz}^2\big) ,
\label{eq:ds-resp}
\end{align}

where every field is at the position $\mathbf r'$ and $s_+\equiv s_{xx}-s_{yy}$. There is a small difference with Blazek in that here we have written explicitly that $C_{1\delta}=b_SC_1$, instead of leaving this as a free parameter, which can absorb any response of the galaxy shape field to the product of the density and tidal fields. 

Now we consider the expectation of Eq.~\eqref{eq:ds-resp} for a given value of $\nu$. Within linear theory, the density, tidal and peak height are jointly Gaussian random fields with zero mean\cite{BBKS1986,vandeWeygaert1996}. For jointly Gaussian random fields we use the following relations for their conditional expectations: ${\rm E}[X|\nu]=\langle X\nu\rangle\,\nu$ and ${\rm E}[XY|\nu]=\langle XY\rangle+\langle X\nu\rangle\langle Y\nu\rangle(\nu^2-1)$\cite{BBKS1986}. To simplify the following equations, we define $c_\delta\equiv\langle\delta\nu\rangle$, $c_+\equiv\langle s_+\nu\rangle$, $c_{zz}\equiv\langle s_{zz}\nu\rangle$, and $c_{xz}\equiv\langle s_{xz}\nu\rangle$. Therefore we arrive at,

\begin{align}
  {\rm E}\big[(1+b_S\delta)\gamma_+\,\big|\,\nu\big]={}&C_1c_+\,\nu+b_SC_1c_\delta c_+\big(\nu^2-1\big)\nonumber\\
  &+C_2\big(c_{xz}^2-c_{zz}c_+\big)\big(\nu^2-1\big) .
\label{eq:ds-cond}
\end{align}

The linear response is proportional to an odd term of $\nu$, whereas the quadratic response is proportional to $\nu^2-1$, meaning that as the peak height goes from negative to positive the sign of the linear response changes whereas the quadratic response keeps the same sign. We then average Eq.~\eqref{eq:ds-cond} over all positions within the quintile, using $\langle\nu\rangle_q=+1.4$ and $-1.4$ for $q_h$ and $q_l$, and $\langle\nu^2-1\rangle_q=1.178$ for both quintiles. Therefore, 

\begin{align}
  \xi_{q+}={}&\pm1.4\,C_1c_++1.178\,b_SC_1c_\delta c_+\nonumber\\
  &+1.178\,C_2\big(c_{xz}^2-c_{zz}c_+\big) ,
\label{eq:ds-model}
\end{align}

where $c_+$, $c_{\delta}$, $c_{xz}$ and $c_{zz}$ are calculated as

\begin{equation}
  c_X(\mathbf r'-\mathbf r)=\frac{b_g}{\sigma_c}\int\!\frac{{\rm d}^3k}{(2\pi)^3}\,W(\mathbf k)\,P_{\delta}(k)\,K_X(\mathbf k)\,e^{{\rm i}\mathbf k\cdot(\mathbf r'-\mathbf r)} ,
\label{eq:ds-cX}
\end{equation}

where we need to account for the smoothing of the galaxy density field with the window function, $W(\mathbf k)=\exp[-\tfrac12k_\perp^2\sigma_\perp^2-\tfrac12k_z^2(\sigma_\parallel^2+\sigma_{\Delta z}^2)]$, which relates $\nu$ to the true matter field. This takes into account the smoothing that we apply to the Gaussian field in order to estimate the peak-height, $\sigma_\perp=5\,h^{-1}$\,Mpc and $\sigma_\parallel=30\,h^{-1}$\,Mpc, as well as the scatter of the photometric redshifts $\sigma_{\Delta z}=21.3\,h^{-1}$\,Mpc. $\sigma_c^2=b_g^2\int{\rm d}^3k\,W^2(\mathbf k)\,P_\delta(k)/(2\pi)^3$ is the variance of the smoothed galaxy density. The kernel $K_X(\mathbf k)$ takes the values $K_\delta=1$, $K_+=(k_x^2-k_y^2)/k^2$, $K_{zz}=k_z^2/k^2-1/3$ and $K_{xz}=k_xk_z/k^2$ for each of $c_{\delta}$, $c_+$, $c_{zz}$ and $c_{xz}$.

\subsection*{Alignment in bins of stellar mass}\label{sec:massbins} 

In Fig.~\ref{fig:mass} we presented the alignment amplitudes as a function of stellar mass. The shape samples in each redshift bin are divided into five stellar mass bins of width 0.5 $\log_{10}(M_\star/M_\odot)$ from 9 to 11.5, and one bin for all galaxies above 11.5. 

It is important to note that our shape samples are not complete in stellar mass. As our sample is flux limited we will miss many low mass galaxies. Additionally our resolution cut, ${\rm res}>1$ (a half-light radius larger than the full width at half maximum of the point-spread function), means that at fixed mass we preferentially select large galaxies. This second point is very important. For different sized galaxies the Tractor shape information will come from different physical scales within the galaxy. Therefore we cannot tell from these results alone whether the mass trends we see arise from the stellar mass of the shape galaxies, or from measuring their shapes at different radii. 

\subsection*{Model fitting and covariance matrices}\label{sec:fitting}

\begin{figure*}[tbp]
  \centering
  \placefig{\textwidth}{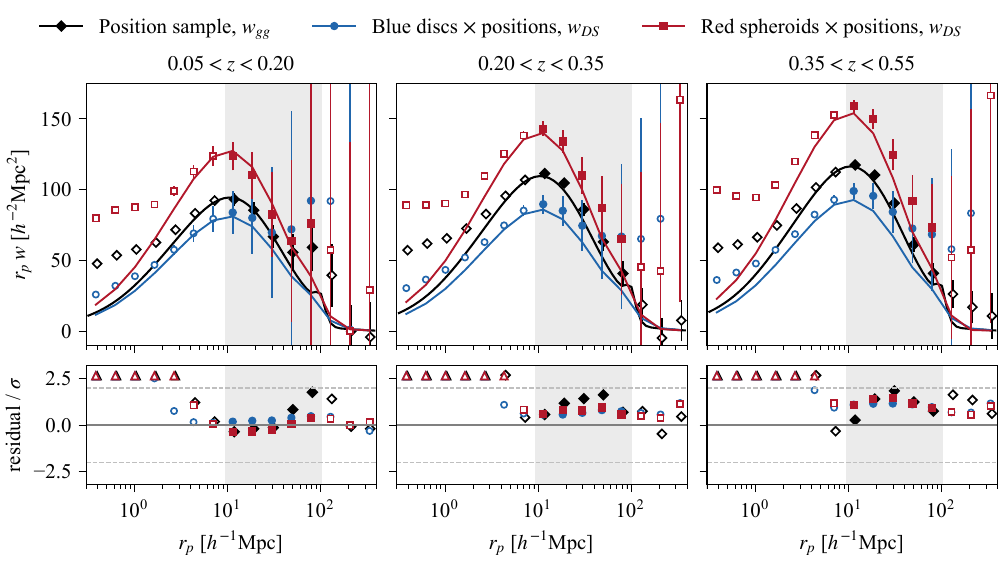}{Clustering of the position sample with the linear-bias fit}
  \caption{\textbf{The galaxy clustering of the position and shape samples follow linear theory.}
Black diamonds: auto-correlations $r_p w_{gg}$ of the galaxy position sample for each redshift bin. Red squares: the cross-correlation between the galaxy position sample and the red spheroid shape sample positions. Blue circles: the cross-correlation between the galaxy position sample and the blue disc sample positions. The solid lines show the best fit for the linear galaxy-bias model. Lower panels: residuals between the observation and the model, divided by the jackknife errors. The triangles mark disagreements beyond $3\sigma$. The small-scale excess is due to non-linear galaxy clustering.}
  \label{fig:wgg}
\end{figure*}

Throughout the paper, each model is fitted between $10\le r_p\le100\,h^{-1}$\,Mpc with five logarithmic bins. The lower limit is a conservative choice. Our models assume linear galaxy bias and perturbative alignment models, so they are only valid when the density field is close to linear. We can test where this breaks down with the clustering of our own samples (Extended Data Fig.~\ref{fig:wgg}). The position sample and the blue disc and red spheroid shape samples agree with linear bias to within $1.2\sigma$ at $r_p\approx7\,h^{-1}$\,Mpc, but lie 1--6\% above it (up to $4\sigma$) at $4\,h^{-1}$\,Mpc and 5--15\% above it (up to $11\sigma$) at $3\,h^{-1}$\,Mpc. Past direct measurements of galaxy alignments have fitted the non-linear linear alignment model above $6\,h^{-1}$\,Mpc\cite{Joachimi2011,Samuroff2023}.

\begin{figure*}[tbp]
  \centering
  \placefig{\textwidth}{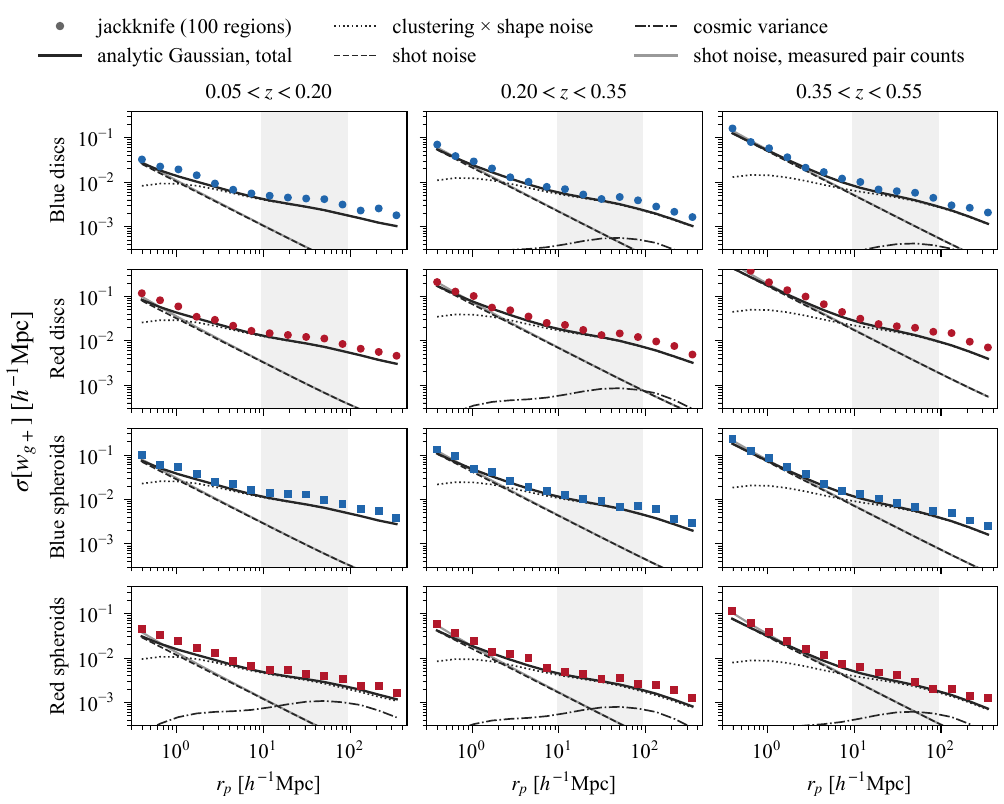}{Diagonal of the w_g+ covariance, jackknife against the analytic terms}
  \caption{\textbf{The jackknife errors are larger than the analytic Gaussian errors on every scale.}
  $\sigma[w_{g+}]$ for each sample (rows) and redshift bin (columns), as in Fig.~\ref{fig:wgp}. Points: the
  100-region jackknife used in every fit. Black: the analytic Gaussian covariance (total, solid) and its terms,
  clustering $\times$ shape noise (dotted), shot noise (dashed) and cosmic variance (dash-dotted). Grey line: the shot
  noise computed with the measured pair counts, which includes the clustering of tracers around the shape galaxies.
  Grey band: fitted scales.}
  \label{fig:covariance}
\end{figure*}

We limit the fits to $r_p\le100\,h^{-1}$\,Mpc because on larger scales it is more difficult to estimate the errors when the size of the jackknife regions approaches the scales that we measure. The jackknife regions are approximately $14^\circ$ across, which is 37--140, 140--240 and 240--350\,$h^{-1}$\,Mpc over the three redshift bins. In the lowest redshift bin the largest fitted scales are already comparable to the size of a region, and its errors on those scales are the least reliable; the jackknife errors there are larger than the analytic Gaussian errors (Extended Data Fig.~\ref{fig:covariance}).

The measurements are then fit using a Gaussian likelihood, using the covariance estimated from the 100 jackknife regions (\nameref{sec:corr}). We compute posteriors using the code emcee\cite{ForemanMackey2013}. We use large flat priors for each parameter:  $-20<A_1<20$, $-100<A_{1\delta}<100$, $-200<A_2<200$ and $0.5<b_g<4$.  In Extended Data Fig.~\ref{fig:covariance} we compare the diagonal of our jackknife covariance against the analytic Gaussian covariance for projected correlation functions, derived by \refcite{ChisariDvorkin2013} and applied to simulated intrinsic alignment measurements by \refcite{Samuroff2021}. We find that our jackknife errors are consistently larger than the theoretical prediction. We therefore adopt the jackknife covariance errors throughout as we consider this to be the more conservative choice. 

\subsection*{Quantifying contamination from weak lensing}\label{sec:lensing}

Weak gravitational lensing of the shape sample by the position sample (galaxy--galaxy lensing) would also produce a tangential shear signal for the blue discs, as we see in the observations. Fig.~\ref{fig:shape} includes the predicted lensing signal (the lensing term $w_{gG}$, following \refcite{Georgiou:2025pca}). As the shape and position samples are at the same redshift, the predicted contamination of our intrinsic alignment signal by weak lensing is very small: 3\% of the blue disc correlation function at $0.20<z<0.35$. Given the redshift distributions of the shape and position samples, to explain the observed signal with gravitational lensing, every galaxy of the position sample would need a halo mass of $\sim\!10^{15.7}\,h^{-1}M_\odot$. In other words every position galaxy would have to be within the most massive clusters in the Universe, which is clearly an unreasonable proposition.

\begin{figure*}[tbp]
  \centering
  \placefig{\textwidth}{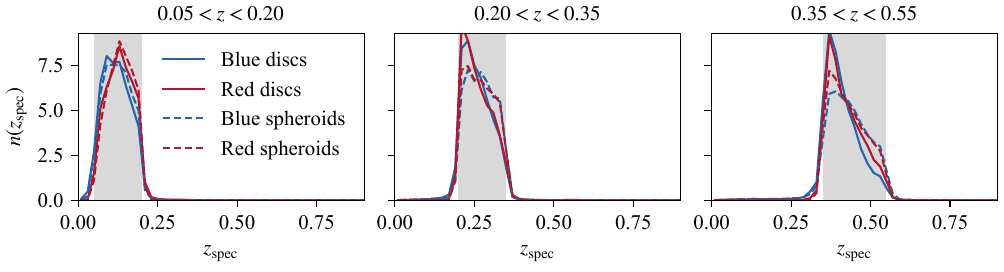}{Photometric against spectroscopic redshift of the shape samples}
  \caption{\textbf{The LSx redshifts of the shape samples are precise, with few catastrophic outliers.}
  Each panel shows the redshift distribution for each of the shape samples with a spectroscopic match, where the redshift distribution is shown using the matched spectroscopic redshift. }
  \label{fig:photoz}
\end{figure*}

\begin{figure*}[tbp]
  \centering
  \placefig{\textwidth}{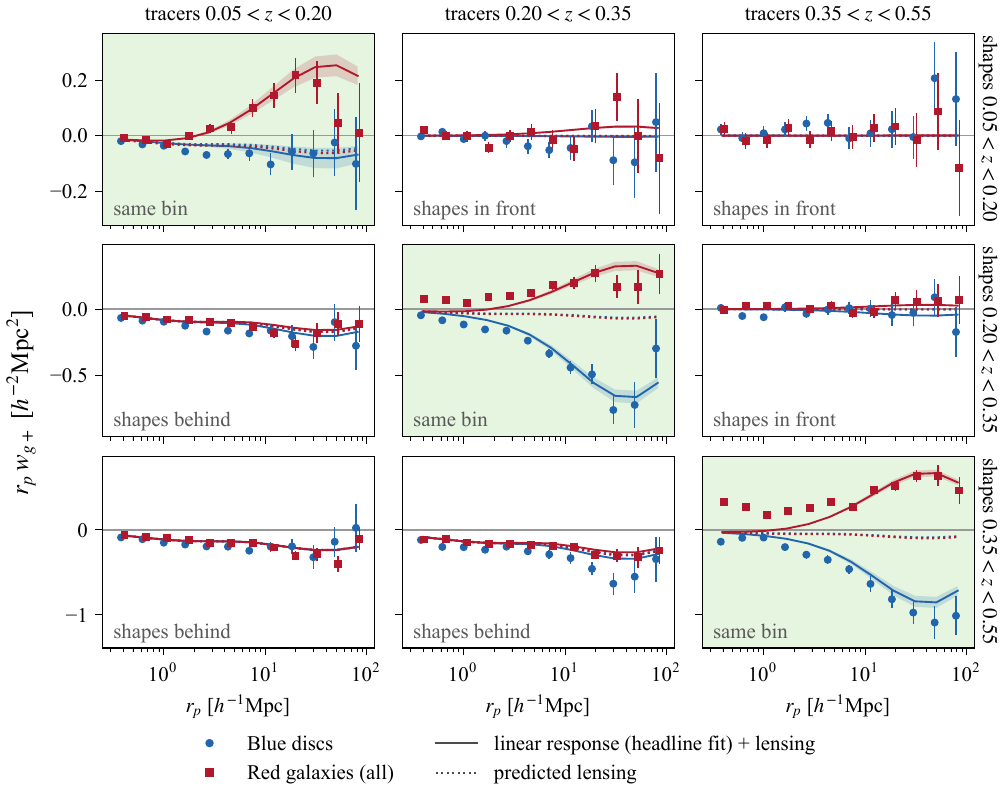}{The alignment between redshift bins}
  \caption{\textbf{There is no evidence of significant lensing contamination. } $r_p w_{g+}$ measurements for every pair of position and shape tracers across different redshift bins (there are no line-of-sight distance limits for these correlation functions). The green panels show the shape-position correlations for galaxies at the same redshift, where we see significant non-zero correlation function measurements, which are consistent with the previous alignment amplitude measurements. The panels with shapes in front of position tracers are consistent with null-detections, as anticipated from precise redshift measurements with low-outlier rates. The panels with shapes behind the position tracers show small non-zero correlations, consistent with the prediction from gravitational lensing.}
  \label{fig:crossz}
\end{figure*}

However we could understimate the lensing prediction if our redshift distributions are incorrect. This is unlikely given the low outlier rates, 0.13--0.52\% (Extended Data Fig.~\ref{fig:photoz}), and the small redshift errors, $\sigma_{\rm NMAD}=0.0055$--0.0077, of the quality sample in our three redshift bins. We test this independently of the spectroscopic validation by cross-correlating the position and shape samples of within redshift bins. In Extended Data Fig.~\ref{fig:crossz} we show these $w_{g+}$ correlations without the condition $|\Pi_{\rm obs}|<\Pi_{\max}$, so that all position--shape pairs are used. From lensing and incorrect redshift distributions, we would see a significant signal in every combination of bins. If the redshift distributions are correct, we expect a small lensing signal where the position sample is in front of the shape sample, no signal where the shapes are in front of the positions, and a strong intrinsic alignment signal where both are in the same redshift bin. We observe the second scenario, therefore our observations are inconsistent with gravitational lensing.

\subsection*{Comparison with published alignment measurements}\label{sec:literature}

\begin{figure*}[tbp]
  \centering
  \placefig{\textwidth}{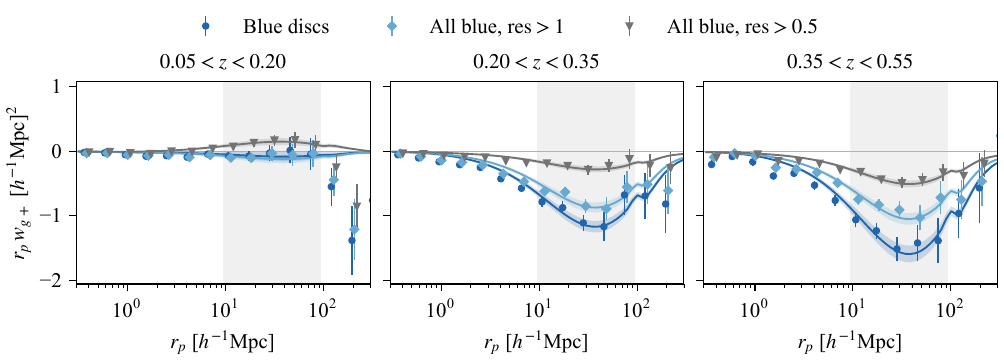}{The alignment of all blue galaxies against blue discs}
  \caption{\textbf{The blue galaxy alignment is smaller than blue disc alignment.} $r_p w_{g+}$ for blue discs (dark blue circles), all well-resolved blue galaxies (light blue diamonds) and all of the blue galaxies with ${\rm res}>0.5$ (grey triangles), in each redshift bin. We see that in each case the intrinsic alignment signal, when detected, is strongest for blue discs followed by all blue galaxies and then the smallest least resolved galaxies (grey triangles).}
  \label{fig:blueall}
\end{figure*}

The large alignment amplitude that we measure for blue discs may appear surprising given past constraints on the alignments of blue galaxies. In Extended Data Fig.~\ref{fig:blueall} we show that the alignment amplitude is much smaller when we select all blue galaxies, without the disc selection: at $0.20<z<0.35$, $A_1=-2.06\pm0.19$ for all blue galaxies with ${\rm res}>1$ and $-0.62\pm0.12$ with ${\rm res}>0.5$, compared with $-2.81\pm0.23$ for the blue discs. This is similar to the result of \refcite{Paviot2026}, where low-mass blue galaxies show no alignment and higher-mass blue galaxies show a $3.3\sigma$ tangential alignment. Our blue galaxy sample is large enough, and its redshifts precise enough, that entire blue galaxy sample without stellar mass or morphological cuts give a significant tangential alignment.

\begin{figure*}[tbp]
  \centering
  \placefig{\textwidth}{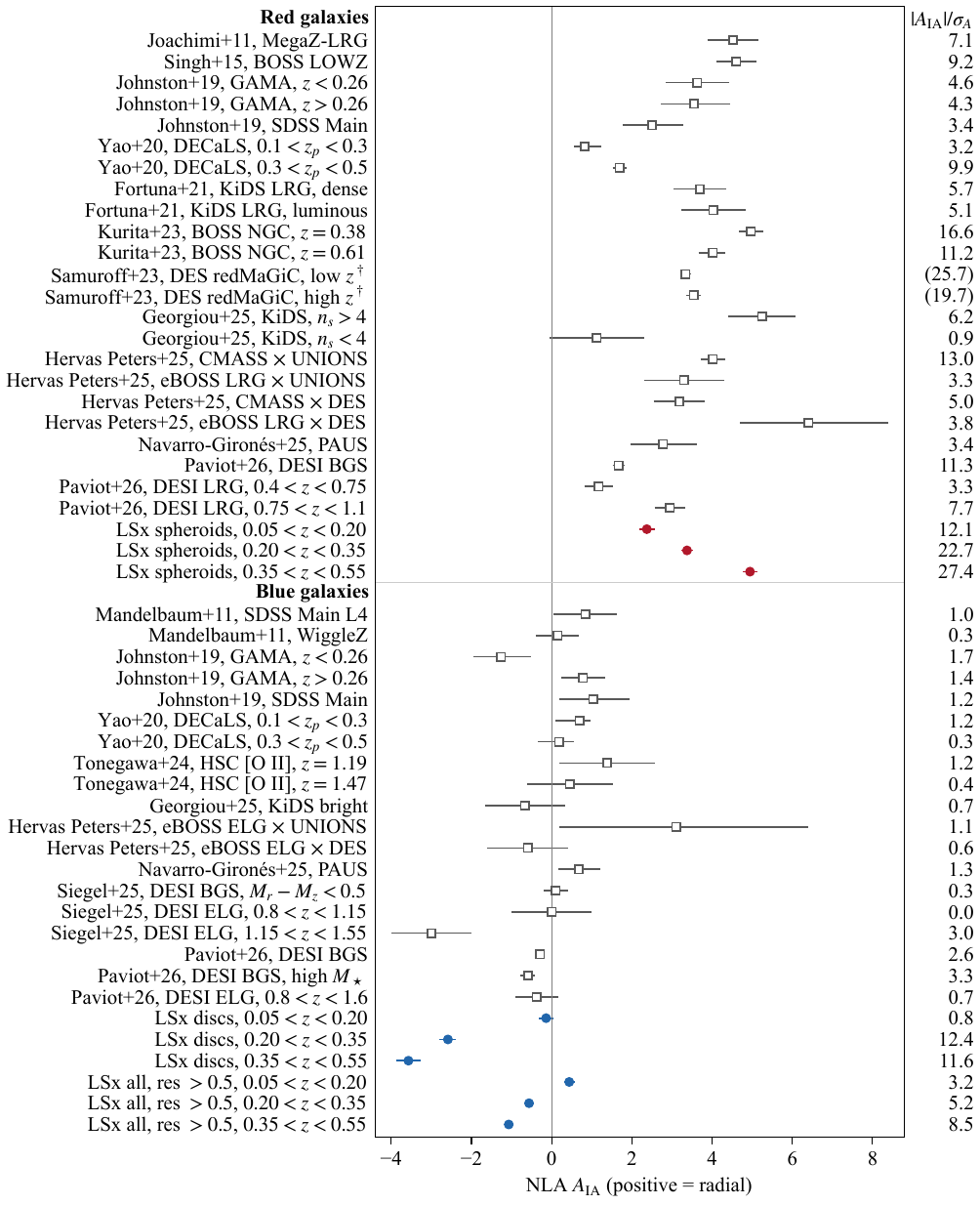}{Comparison with DESI and published alignment amplitudes}
  \caption{ \textbf{Our results are consistent with past results for similar galaxy types.}
  Alignment amplitudes of the NLA model for red galaxies (top) and blue galaxies (bottom): the published measurements (open squares)\cite{Joachimi2011,Mandelbaum2011,Singh2015,Johnston:2018nfi,Yao:2020jpj,Fortuna2021,Kurita:2023qku,Samuroff2023,Tonegawa:2024xuw,HervasPeters2025,Georgiou:2025pca,DESI:2025fda,Navarro-Girones:2025tyh,Paviot2026} and this work (filled circles). The right-hand column gives the significance of each measurement, $|A_{\rm IA}|/\sigma_A$. The 95\% limits of \refcite{Mandelbaum2011,Tonegawa:2024xuw} are divided by 1.96 to approximate the $1-\sigma$ constraints of the other analyses. The two redMaGiC samples from \refcite{Samuroff2023} ($\dagger$) are shown as published, however, \refcite{HervasPeters2025} found that the covariance used in that work omits the term from clustering power and shape noise. We therefore expect their significances to be overestimated by a factor of around 1.5.}
  \label{fig:literature}
\end{figure*}

In Extended Data Fig.~\ref{fig:literature} we compare our results with published measurements. As past work has generally used the NLA model, we refit our measurements with the NLA model for this comparison. Our blue discs show a significant and large amplitude tangential alignment, in contrast to previous measurements. When we instead select all blue galaxies with ${\rm res}>0.5$, we still detect an alignment in every redshift bin, but the amplitudes are comparable with those of past measurements. This detection is due to the improved statistical power of this sample compared to past observations. 

For red spheroids our results are consistent with the literature. Red spheroids show significant radial alignments, which increase with galaxy stellar mass. Our measurement at $0.35<z<0.55$ ($27.4\sigma$ by $|A_{\rm IA}|/\sigma_A$) is the most significant detection of radial alignment of red galaxies to date. 

We also measured shape-shape correlations $w_{++}$ and $w_{\times \times}$ for each of our samples. None of these correlations is significantly different from zero, but they are consistent with the amplitudes predicted from the observed $w_{g+}$ with the linear response model.

\subsection*{Consistent results using DES Y6 shapes}\label{sec:des}

The Tractor shapes are corrected for the point-spread function but they are not calibrated for weak lensing shear, therefore the multiplicative bias is unknown. A multiplicative bias rescales the measured $A_1$, however it cannot produce a correlation that does not otherwise exist or even change the significance of the detected correlation. This is therefore not a major concern for this work, the aim of which is primarily to improve our understanding of galaxy intrinsic alignments rather than to provide alignment priors for weak lensing experiments. The Tractor shapes of the Legacy Imaging Surveys have not been extensively tested. The Dark Energy Survey Year 6 (DES Y6) shape catalogue, was built specifically for weak lensing and has been thoroughly validated\cite{Yamamoto2025}. We therefore repeat our measurements with DES Y6 shapes.

We match LSx positions to DES Y6 shapes within $1''$. Of the 3.92 million galaxies of the LSx shape sample within the 4750\,deg$^2$ DES footprint, 3.02 million have a DES Y6 match. The majority of the unmatched galaxies are large and bright; DES excludes large bright galaxies as a large fraction of their large bright galaxies are imaging systematics. In each sample and redshift bin we divide the DES shapes by the mean metacalibration shear response. This response is for calibrating weak lensing shear, rather than intrinsic galaxy shapes so it is unclear that this correction is correct. The effect is simply to rescale the alignment amplitude.

\begin{figure*}[tbp]
  \centering
  \placefig{\textwidth}{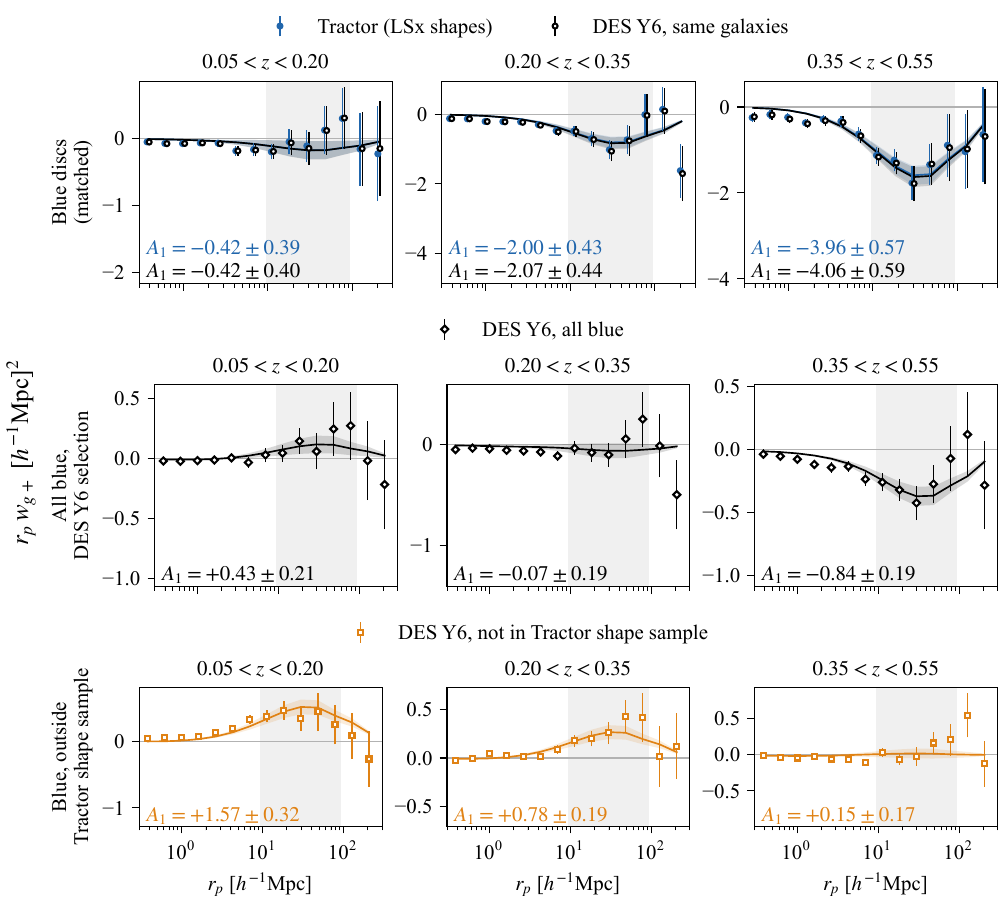}{$w_{g+}$ from Tractor shapes against DES Y6 shapes}
  \caption{\textbf{An independent shape measurement gives the same alignments.} $r_p w_{g+}$ measured with DES Y6 galaxy shapes. Top: blue discs matched between DES Y6 and the Tractor, with Tractor (filled) and DES Y6 (open) shapes of the same galaxies; the amplitudes are consistent in each bin. Middle: blue galaxies passing the DES Y6 shape cuts, without the morphological or size cuts of the Tractor sample. The alignment amplitude is much smaller, but still significant in the highest redshift bin. Bottom: the galaxies of the middle row that are not in the Tractor shape sample. These galaxies show significant radial alignments in the low and intermediate redshift bins.}
  \label{fig:desshapes}
\end{figure*}

The results are shown in Extended Data Fig.~\ref{fig:desshapes}. For the same blue disc galaxies, the DES Y6 and Tractor shapes give consistent amplitudes $\Delta A_1=-0.07\pm0.04$ and $-0.14\pm0.07$ at $0.20<z<0.35$ and $0.35<z<0.55$. With the DES Y6 shapes the tangential alignment of the blue discs is detected at $4.7\sigma$ and $6.9\sigma$ in these bins; the detection significance is lower than in the main analysis as the survey footprint of DES is smaller 4750\,deg$^2$ rather than 19,312\,deg$^2$ of the LSx catalogue. For all red galaxies (not shown in the Figure for brevity) we also find significant radial alignments with DES Y6 shapes, however the amplitudes are 16--29\% lower than with the Tractor shapes of the same galaxies ($A_1=+1.39\pm0.40$ against $+1.94\pm0.36$ at $0.20<z<0.35$, and $+3.75\pm0.47$ against $+4.44\pm0.40$ at $0.35<z<0.55$). It has been shown previously that the alignment amplitude is sensitive to the shape measurement method\cite{SinghMandelbaum2016,Georgiou2019}.

We also consider a broader DES Y6 shape sample, where the only selection are the DES Y6 shape selection and the LSx colour cut. Therefore no morphological or resolution cuts. This sample has a weaker alignment than the blue discs, $A_1=+0.43\pm0.21$, $-0.07\pm0.19$ and $-0.84\pm0.19$ in the three redshift bins, against $-0.42\pm0.40$, $-2.07\pm0.44$ and $-4.06\pm0.59$ for the matched blue discs. The main difference is that this sample contains galaxies of a smaller angular size. Further we consider the galaxies that our in this sample, which are not within the Tractor sample. These galaxies align radially, with $A_1=+1.57\pm0.32$ and $+0.78\pm0.19$ in the two lower redshift bins, and are consistent with no alignment $+0.15\pm0.17$ in the highest redshift bin. The alignment of blue galaxies is therefore strongly dependent upon not only morphology but size and resolution. Small blue galaxies align radially, and large blue discs align tangentially.

\subsection*{Use of a large language model}\label{sec:llm}

Claude (Anthropic), was used to write code for this research project and language editing of the manuscript. The authors take full responsibility for the contents of this paper.

\section*{Data availability}
The DESI Legacy Imaging Surveys, SPHEREx, DES, DESI DR1, SDSS, GAMA, 6dFGS and 2dFGRS data used in this work are publicly available from the respective surveys. The measured correlation functions and the data underlying each figure are available on request.

\section*{Code availability}
The correlation functions in this work were made with TreeCorr\cite{Jarvis2004}, the model predictions were computed with CAMB\cite{Lewis2000} and FAST-PT\cite{Fang2017}, the posteriors were sampled using emcee\cite{ForemanMackey2013} and the stellar masses estimated with CIGALE\cite{Boquien2019}, all of which are publicly available. The analysis and plotting code is available on request.

\ifauthoryear\bibliographystyle{plainnat}\else\bibliographystyle{naturemag}\fi
\bibliography{references}

\section*{Acknowledgements}

Calum Murray would like to thank Sandrine Codis, Olivier Doré, Jamie Bock, Richard M. Feder, and Zhaoyu Huai for useful discussions about this work. This publication makes use of data products from SPHEREx, a joint project of the Jet Propulsion Laboratory and the California Institute of Technology, funded by the National Aeronautics and Space Administration. The DESI Legacy Imaging Surveys consist of three individual and complementary projects: the Dark Energy Camera Legacy Survey (DECaLS), the Beijing-Arizona Sky Survey (BASS), and the Mayall z-band Legacy Survey (MzLS); the complete acknowledgments can be found at \url{https://www.legacysurvey.org/acknowledgment/}. This project used data from the Dark Energy Survey (DES) and the Dark Energy Spectroscopic Instrument (DESI), whose full acknowledgments are at \url{https://www.darkenergysurvey.org/the-des-project/data-access/} and \url{https://data.desi.lbl.gov/doc/acknowledgments/}. Funding for SDSS-IV has been provided by the Alfred P. Sloan Foundation, the U.S. Department of Energy Office of Science, and the Participating Institutions. GAMA is a joint European-Australasian project based around a spectroscopic campaign using the Anglo-Australian Telescope (\url{https://www.gama-survey.org/}). We also use redshifts from the 6dF Galaxy Survey and the 2dF Galaxy Redshift Survey.

\clearpage

\begin{table*}[htbp]
  \centering
  \caption{\textbf{Galaxies in each shape sample.} Number of galaxies in each redshift bin.}
  \label{tab:samples}
  \small
  \begin{tabular}{lrrr}
    \hline
     & $0.05<z<0.20$ & $0.20<z<0.35$ & $0.35<z<0.55$ \\
    \hline
    Blue discs      & 2,797,143 & 1,844,256 & 862,574 \\
    Red discs       &   529,710 &   326,499 & 121,670 \\
    Blue spheroids  &   270,776 &   332,238 & 426,514 \\
    Red spheroids   & 1,185,148 & 1,598,384 & 1,428,209 \\
    \hline
  \end{tabular}
\end{table*}

\begin{table*}[tbp]
  \centering
  \caption{\textbf{The alignment amplitudes for every sample with different models.} We fit each measurement with six different models which are different combinations of the three parameters $A_1$, $A_{1\delta}$ and $A_2$, with linear galaxy-biasing used throughout. Unless otherwise stated all other parameters are set to zero. Model (1) $A_1$ free; model (2) $A_1$ free with the density weighting fixed at $A_{1\delta}=b_SA_1$; model (3) $A_1$ and $A_{1\delta}$ free; model (4) $A_2$ free; model (5) $A_1$ and $A_2$ free with $A_{1\delta}=b_SA_1$; model (6) $A_1$, $A_{1\delta}$ and $A_2$ all free. The density-sample bias $b_D$ is fitted jointly, constrained by $w_{gg}$, and the lensing term $w_{gG}$ is fixed at its prediction. Amplitudes are posterior means and standard deviations. $b_S$ is the measured galaxy bias for each shape sample. $|A_1|/\sigma_A$ is the detection significance from model (1). $\chi^2_\nu$ is evaluated at the best fit over the five bins with $10\le r_p\le100\,h^{-1}$\,Mpc, using the jackknife covariance with the Hartlap correction, with $\nu$ the number of bins minus the number of free amplitudes; the PTE is the probability of a $\chi^2$ at least this large for $\nu$ degrees of freedom. Rows with $|A_1|/\sigma_A\ge5$ are shaded: green where the model is acceptable (PTE $\ge0.05$) and red where it is not.
  }
  \label{tab:fits}
  \small
  \small\setlength{\tabcolsep}{4pt}\definecolor{modelfail}{HTML}{EFA3A3}\definecolor{modelpass}{HTML}{DCF0D8}
\begin{tabular}{llllrrrr}
\hline
 & & & & \multicolumn{2}{c}{(1) $A_1$} & \multicolumn{2}{c}{(2) $A_1$, $A_{1\delta}=b_SA_1$} \\
Sample & $z$ & $b_S$ & $|A_1|/\sigma_A$ & $A_1$ & $\chi^2_4$ (PTE) & $A_1$ & $\chi^2_4$ (PTE) \\
\hline
Blue discs & 0.05--0.20 & $0.96$ & 0.7 & $-0.15\pm0.21$ & 0.3 (0.99) & $-0.11\pm0.13$ & 0.2 (0.99) \\
 & 0.20--0.35 & $0.97$ & 12.4 & \cellcolor{modelpass}$-2.81\pm0.23$ & \cellcolor{modelpass}2.9 (0.58) & \cellcolor{modelpass}$-2.00\pm0.16$ & \cellcolor{modelpass}4.5 (0.34) \\
 & 0.35--0.55 & $1.12$ & 11.6 & \cellcolor{modelpass}$-3.79\pm0.33$ & \cellcolor{modelpass}2.3 (0.68) & \cellcolor{modelpass}$-2.75\pm0.24$ & \cellcolor{modelpass}5.6 (0.23) \\
\hline
Red discs & 0.05--0.20 & $1.26$ & 0.6 & $+0.40\pm0.63$ & 0.8 (0.94) & $+0.23\pm0.37$ & 0.8 (0.93) \\
 & 0.20--0.35 & $1.25$ & 5.9 & \cellcolor{modelpass}$-4.32\pm0.74$ & \cellcolor{modelpass}2.8 (0.59) & \cellcolor{modelpass}$-2.62\pm0.48$ & \cellcolor{modelpass}7.8 (0.10) \\
 & 0.35--0.55 & $1.53$ & 1.4 & $-1.34\pm0.98$ & 0.9 (0.93) & $-0.72\pm0.63$ & 1.4 (0.84) \\
\hline
Blue spheroids & 0.05--0.20 & $0.96$ & 1.1 & $-0.70\pm0.61$ & 2.9 (0.58) & $-0.56\pm0.39$ & 2.1 (0.71) \\
 & 0.20--0.35 & $1.11$ & 1.3 & $+0.52\pm0.40$ & 0.4 (0.99) & $+0.34\pm0.28$ & 0.6 (0.97) \\
 & 0.35--0.55 & $1.27$ & 0.2 & $+0.07\pm0.43$ & 3.0 (0.57) & $-0.02\pm0.30$ & 3.0 (0.56) \\
\hline
Red spheroids & 0.05--0.20 & $1.38$ & 12.2 & \cellcolor{modelpass}$+2.69\pm0.22$ & \cellcolor{modelpass}1.1 (0.90) & \cellcolor{modelpass}$+1.45\pm0.12$ & \cellcolor{modelpass}8.1 (0.09) \\
 & 0.20--0.35 & $1.51$ & 22.9 & \cellcolor{modelpass}$+3.72\pm0.16$ & \cellcolor{modelpass}2.8 (0.59) & \cellcolor{modelfail}$+2.13\pm0.10$ & \cellcolor{modelfail}33.4 (0.00) \\
 & 0.35--0.55 & $1.82$ & 27.1 & \cellcolor{modelfail}$+5.25\pm0.19$ & \cellcolor{modelfail}17.9 (0.00) & \cellcolor{modelfail}$+3.26\pm0.12$ & \cellcolor{modelfail}40.6 (0.00) \\
\hline
\end{tabular}
\par\medskip
\begin{tabular}{llrrrrr}
\hline
 & & \multicolumn{3}{c}{(3) $A_1$, $A_{1\delta}$} & \multicolumn{2}{c}{(4) $A_2$} \\
Sample & $z$ & $A_1$ & $A_{1\delta}$ & $\chi^2_3$ (PTE) & $A_2$ & $\chi^2_4$ (PTE) \\
\hline
Blue discs & 0.05--0.20 & $-0.02\pm0.38$ & $-0.3\pm0.6$ & 0.2 (0.98) & $-0.3\pm0.3$ & 0.1 (1.00) \\
 & 0.20--0.35 & \cellcolor{modelpass}$-2.54\pm0.38$ & \cellcolor{modelpass}$-0.7\pm0.8$ & \cellcolor{modelpass}2.1 (0.56) & \cellcolor{modelfail}$-4.3\pm0.4$ & \cellcolor{modelfail}31.0 (0.00) \\
 & 0.35--0.55 & \cellcolor{modelpass}$-3.52\pm0.46$ & \cellcolor{modelpass}$-0.9\pm1.1$ & \cellcolor{modelpass}1.7 (0.63) & \cellcolor{modelfail}$-6.0\pm0.6$ & \cellcolor{modelfail}39.1 (0.00) \\
\hline
Red discs & 0.05--0.20 & $+0.36\pm0.91$ & $+0.1\pm1.4$ & 0.8 (0.85) & $+0.4\pm0.9$ & 1.0 (0.91) \\
 & 0.20--0.35 & \cellcolor{modelpass}$-5.48\pm1.22$ & \cellcolor{modelpass}$+3.1\pm2.5$ & \cellcolor{modelpass}1.4 (0.72) & \cellcolor{modelfail}$-5.6\pm1.2$ & \cellcolor{modelfail}16.4 (0.00) \\
 & 0.35--0.55 & $-2.14\pm1.57$ & $+2.5\pm3.8$ & 0.4 (0.93) & $-1.4\pm1.7$ & 2.1 (0.72) \\
\hline
Blue spheroids & 0.05--0.20 & $+0.32\pm1.00$ & $-1.9\pm1.4$ & 1.2 (0.74) & $-1.3\pm0.8$ & 1.5 (0.83) \\
 & 0.20--0.35 & $+0.61\pm0.60$ & $-0.2\pm1.2$ & 0.3 (0.96) & $+0.6\pm0.7$ & 1.1 (0.89) \\
 & 0.35--0.55 & $+0.40\pm0.64$ & $-1.2\pm1.5$ & 2.4 (0.50) & $-0.3\pm0.8$ & 2.9 (0.58) \\
\hline
Red spheroids & 0.05--0.20 & \cellcolor{modelpass}$+2.43\pm0.37$ & \cellcolor{modelpass}$+0.5\pm0.6$ & \cellcolor{modelpass}0.3 (0.97) & \cellcolor{modelfail}$+3.3\pm0.3$ & \cellcolor{modelfail}30.9 (0.00) \\
 & 0.20--0.35 & \cellcolor{modelpass}$+3.53\pm0.27$ & \cellcolor{modelpass}$+0.4\pm0.5$ & \cellcolor{modelpass}2.1 (0.55) & \cellcolor{modelfail}$+5.1\pm0.2$ & \cellcolor{modelfail}119.4 (0.00) \\
 & 0.35--0.55 & \cellcolor{modelfail}$+4.65\pm0.28$ & \cellcolor{modelfail}$+2.1\pm0.7$ & \cellcolor{modelfail}8.9 (0.03) & \cellcolor{modelfail}$+8.8\pm0.4$ & \cellcolor{modelfail}216.6 (0.00) \\
\hline
\end{tabular}
\par\medskip
\begin{tabular}{llrrrrrrr}
\hline
 & & \multicolumn{3}{c}{(5) $A_1$, $A_{1\delta}=b_SA_1$, $A_2$} & \multicolumn{4}{c}{(6) $A_1$, $A_{1\delta}$, $A_2$} \\
Sample & $z$ & $A_1$ & $A_2$ & $\chi^2_3$ (PTE) & $A_1$ & $A_{1\delta}$ & $A_2$ & $\chi^2_2$ (PTE) \\
\hline
Blue discs & 0.05--0.20 & $+0.07\pm0.56$ & $-0.4\pm1.2$ & 0.1 (0.99) & $+0.20\pm0.85$ & $+0.9\pm4.2$ & $-1.3\pm4.6$ & 0.1 (0.95) \\
 & 0.20--0.35 & \cellcolor{modelpass}$-2.90\pm0.53$ & \cellcolor{modelpass}$+2.2\pm1.2$ & \cellcolor{modelpass}1.4 (0.71) & \cellcolor{modelpass}$-3.24\pm0.80$ & \cellcolor{modelpass}$-5.7\pm4.9$ & \cellcolor{modelpass}$+5.0\pm4.9$ & \cellcolor{modelpass}1.0 (0.61) \\
 & 0.35--0.55 & \cellcolor{modelpass}$-4.06\pm0.66$ & \cellcolor{modelpass}$+3.5\pm1.7$ & \cellcolor{modelpass}1.1 (0.77) & \cellcolor{modelpass}$-4.39\pm1.11$ & \cellcolor{modelpass}$-7.5\pm7.8$ & \cellcolor{modelpass}$+6.2\pm7.2$ & \cellcolor{modelpass}1.0 (0.61) \\
\hline
Red discs & 0.05--0.20 & $+0.67\pm1.35$ & $-1.1\pm3.1$ & 0.7 (0.87) & $+2.19\pm2.34$ & $+10.9\pm13.0$ & $-11.7\pm13.9$ & 0.1 (0.95) \\
 & 0.20--0.35 & \cellcolor{modelpass}$-7.05\pm1.83$ & \cellcolor{modelpass}$+11.6\pm4.6$ & \cellcolor{modelpass}1.6 (0.66) & \cellcolor{modelpass}$-6.02\pm2.60$ & \cellcolor{modelpass}$-0.4\pm15.7$ & \cellcolor{modelpass}$+3.5\pm15.6$ & \cellcolor{modelpass}1.3 (0.52) \\
 & 0.35--0.55 & $-3.12\pm2.34$ & $+6.9\pm6.4$ & 0.3 (0.96) & $-3.28\pm3.65$ & $-6.5\pm26.3$ & $+8.3\pm24.0$ & 0.3 (0.86) \\
\hline
Blue spheroids & 0.05--0.20 & $+0.70\pm1.35$ & $-2.7\pm2.8$ & 1.2 (0.76) & $+0.70\pm1.97$ & $+0.7\pm10.9$ & $-2.8\pm11.6$ & 1.2 (0.55) \\
 & 0.20--0.35 & $+0.73\pm0.81$ & $-1.0\pm2.0$ & 0.3 (0.96) & $+0.65\pm1.23$ & $+0.1\pm8.8$ & $-0.3\pm8.5$ & 0.3 (0.85) \\
 & 0.35--0.55 & $+0.71\pm0.86$ & $-2.0\pm2.3$ & 2.2 (0.53) & $+1.00\pm1.25$ & $+4.2\pm10.0$ & $-4.9\pm9.0$ & 2.1 (0.35) \\
\hline
Red spheroids & 0.05--0.20 & \cellcolor{modelpass}$+2.77\pm0.51$ & \cellcolor{modelpass}$-3.3\pm1.2$ & \cellcolor{modelpass}0.9 (0.82) & \cellcolor{modelpass}$+2.44\pm0.65$ & \cellcolor{modelpass}$+0.6\pm4.1$ & \cellcolor{modelpass}$-0.1\pm4.2$ & \cellcolor{modelpass}0.3 (0.87) \\
 & 0.20--0.35 & \cellcolor{modelpass}$+4.18\pm0.39$ & \cellcolor{modelpass}$-5.5\pm1.0$ & \cellcolor{modelpass}4.1 (0.25) & \cellcolor{modelpass}$+3.55\pm0.60$ & \cellcolor{modelpass}$+0.6\pm4.1$ & \cellcolor{modelpass}$-0.1\pm4.0$ & \cellcolor{modelpass}2.1 (0.35) \\
 & 0.35--0.55 & \cellcolor{modelpass}$+5.51\pm0.38$ & \cellcolor{modelpass}$-7.2\pm1.2$ & \cellcolor{modelpass}0.8 (0.84) & \cellcolor{modelpass}$+5.74\pm0.46$ & \cellcolor{modelpass}$+13.3\pm3.8$ & \cellcolor{modelpass}$-10.0\pm3.4$ & \cellcolor{modelpass}0.1 (0.97) \\
\hline
\end{tabular}

\end{table*}

\end{document}